%% file: VIMOS_paper2_v7_MNRAS.tex
\documentclass[a4paper,fleqn,usenatbib,useAMS]{mnras}
 \usepackage{graphicx}
 \usepackage[T1]{fontenc}
 \usepackage{multirow}
 \usepackage{times}
 \usepackage{amssymb}

 \newcommand{\COzero}{${}^{12}\mathrm{CO}(1-0)$}
 \newcommand{\COtwo}{${}^{12}\mathrm{CO}(2-1)$}
 
 \newcommand{\Ox}{\textsc{[oiii]}}

 \newcommand{\Ha}{\mbox{H$\alpha$}}
 
 \newcommand{\QDeb}{\textsc{qdeblend${}^{\mathrm{3D}}$}}

\begin{document}

\title[Integral field spectroscopy of nearby QSOs II.]{Integral field spectroscopy of nearby QSOs II. The molecular gas content and condition for star formation}

\author[Husemann et al.]{B.~Husemann$^{1,2}$, T.~A.~Davis$^{3}$, K.~Jahnke$^1$, H.~Dannerbauer$^{4,5}$,  T.~Urrutia$^{6}$, J.~Hodge$^{7}$\newauthor\\
$^1$Max-Planck-Institut f\"ur Astronomie, K\"onigstuhl 17, D-69117 Heidelberg, Germany\thanks{husemann@mpia.de}\\
$^2$European Southern Observatory, Karl-Schwarzschild-Str. 2, D-85748 Garching b. M\"unchen, Germany\\
$^3$School of Physics \&\ Astronomy, Cardiff University, Queens Buildings, The Parade, Cardiff, CF24 3AA, UK\\
$^4$Instituto de Astrof\'isica de Canarias (IAC), E-38205 La Laguna, Tenerife, Spain\\
$^5$Universidad de La Laguna, Dpto. Astrof\'isica, E-38206 La Laguna, Tenerife, Spain\\
$^6$Leibniz-Institut f\"ur Astrophysik Potsdam, An der Sternwarte 16, D-14482 Potsdam, Germany\\
$^7$Leiden Observatory, Leiden University, P.O. Box 9513, 2300 RA Leiden, Netherlands
}
\maketitle
\begin{abstract}
We present single-dish \COzero\ and  \COtwo\ observations for 14 low-redshift quasi-stellar objects (QSOs). In combination with optical integral field spectroscopy we study how the cold gas content relates to the star formation rate (SFR) and black hole  accretion rate. \COzero\ is detected in 8 of 14 targets and \COtwo\ is detected in 7 out of 11 cases. The majority of disc-dominated QSOs reveal gas fractions and depletion times well matching normal star forming systems. Two gas-rich major mergers show clear starburst signatures with higher than average gas fractions and shorter depletion times. Bulge-dominated QSO hosts are mainly undetected in \COzero\ which corresponds, on average, to lower gas fractions than in disc-dominated counterparts. Their SFRs however imply shorter than average depletion times and higher star formation efficiencies. Negative QSO feedback through removal of cold gas seems to play a negligible role in our sample. We find a trend between black hole accretion rate and total molecular gas content for disc-dominated QSOs when combined with literature samples. We interpret this as an upper envelope for nuclear activity which is well represented by a scaling relation between the total and circum-nuclear gas reservoir accessible for accretion. Bulge-dominated QSOs significantly differ from that scaling relation and appear uncorrelated with the total molecular gas content. This could be explained either by a more compact gas reservoir, blow out of the gas envelope through outflows, or a different ISM phase composition. 
\end{abstract}

\section{Introduction}
Luminous active galactic nuclei (AGN) are thought to influence the evolution of their host galaxies by destroying, heating or removing the gas from which stars are being formed. AGN feedback is an essential ingredient nowadays for the vast majority of semi-analytical models and numerical simulations to explain the luminosity function and colors of massive galaxies $\geq10^{10}\mathrm{M}_\odot$ \citep[see][for a review]{Somerville:2015}. The mechanical energy of powerful radio jets released by an AGN has been shown to significantly heat the hot gas in galaxy clusters or groups to prevent gas cooling and hence star formation in the central galaxy \citep[see][for a review]{Fabian:2012}. However, this so-called ''radio-mode'' feedback or ''maintenance mode'' usually prevents star formation in sufficiently dense environments in which star formation already ceased. To actively suppress star formation for the majority of massive galaxies may require the so-called ''Quasar-mode'' feedback where the AGN luminosity is thought to expel or heat the cold gas from the galaxy \citep[e.g.][]{Silk:1998,King:2003}. Firm evidence for this mode of AGN feedback is still elusive.

Different approaches to reveal the nature of AGN feedback have been pursued. Direct detection of AGN-driven outflows on kpc scales are reported in the ionized gas \citep[e.g.][]{Greene:2011,Liu:2013b,Harrison:2014,Brusa:2015,Perna:2015,McElroy:2015,Kakkad:2016}, neutral gas \citep[e.g.][]{Rupke:2015,Morganti:2016} and molecular gas phase \citep[e.g.][]{Alatalo:2011,Sturm:2011,Cicone:2014,Feruglio:2015,Morganti:2015,Zschaechner:2016}. Nevertheless, it still debated whether very extended ionized outflows are ubiquitous for all luminous AGN \citep{Husemann:2013a,Karouzos:2016,Villar-Martin:2016,Husemann:2016c}. To assume a causal connection between kpc-scale outflows and the quenching of star formation remain tempting. Recent IFU observations of luminous AGN $z\sim2$ have shown evidence for this ``negative'' feedback given the spatial distribution of the outflows and the sites of star formation as traced by H$\alpha$ \citep[e.g.][]{Cresci:2015,Carniani:2016}. However, the expanding shock front of the outflow may also trigger star formation through the compression of gas leading to ''positive'' AGN feedback \citep[e.g.][]{Croft:2006,Silk:2013,Cresci:2015,Cresci:2015b,Bieri:2016}. 

A large number of studies compared the total star formation rate (SFR), which should reveal the net effect of positive and negative feedback, with the AGN luminosity or outflow characteristics. The results remain inconclusive so far. Recent studies reported either a positive correlation \citep{Netzer:2009,Chen:2013,Delvecchio:2015,Matsuoka:2015a,Xu:2015,Guerkan:2015,Heinis:2016,Dong:2016,Harris:2016} or no correlation \citep{Rosario:2012,Azadi:2015,Stanley:2015,Shimizu:2016}   of SFR with increasing AGN luminosity. Other studies have compared the specific SFR in luminous AGN compared to a control sample of non-AGN galaxies at the respective redshift and report either that the SFR is suppressed \citep{Mullaney:2015,Shimizu:2015,Wylezalek:2016}, equal \citep{Rosario:2013,Xu:2015} or enhanced \citep{Zhang:2016,Bernhard:2016} in luminous AGN hosts. It has been shown that part of the huge discrepancies in the results may be caused by sample selection effects, e.g. binning in SFR or AGN luminosity \citep{Volonteri:2015}, that the FIR-based SFRs are over-estimated in post-starburst galaxies \citep{Hayward:2014}, and/or by a timescale 
discrepancy between the AGN phase and the star formation in the host galaxy \citep{Hickox:2014}. 

The cold molecular gas provides an instantaneous measure of the raw fuel for star formation in galaxies. Hence, it may provide a more direct tracer for the impact of AGN in terms of positive and negative feedback in the galaxies ability to form stars. \citet{Maiolino:1997} performed a large systematic study of \COzero\ in 73 nearby AGN host to estimate their molecular gas reservoir, which turned out  to be consistent with those of non-AGN galaxies. \citet{Saintonge:2012} reported that long gas depletion times in nearby bulge-dominated galaxies are not statistically linked to AGN. However, the low luminosity of these nearby AGN may simply prevent any effect on global host galaxy properties. CO observations of more luminous AGN and Quasi-Stellar Objects (QSOs) are challenging due to their higher redshifts, so that it has been difficult to built up large samples. Over the years several studies have gathered significant data for QSOs at $z<1$  \citep[e.g.][]{Evans:2001,Scoville:2003,Evans:2006,Bertram:2007, Krips:2012, Xia:2012,Villar-Martin:2013, Rodriguez:2014} and $z>2$ \citep[e.g.][]{Carilli:2002,Riechers:2006,Maiolino:2007,Coppin:2008}. Most of these studies report that the QSO host galaxies are rich in molecular gas in contrast to the expectations that ''Quasar-mode'' feedback has erased the gas content.

In \citet[][hereafter Paper I]{Husemann:2014}, we presented integral-field unit (IFU) spectroscopy for a sample of 18 luminous unobscured (type I) QSOs in the redshift range $0.04<z<0.2$. After careful decomposition of the AGN and host galaxy emission, we mapped the location of ionized gas excited by young stars and measured a dust-corrected SFR based on the H$\alpha$ luminosity. We reported that most of our systems are following the main-sequence of star formation irrespective of whether the host galaxies are bulge- or disc-dominated systems. In this paper, we present  \COzero\ and \COtwo\ line follow-up observations with the IRAM 30m telescope for 14 out of 18 QSOs in the sample to infer their total molecular gas content and depletion time scales. Our primary aim is to discriminate whether any changes in the SFR in those luminous AGN are primarily driven by the total gas content or the star formation efficiency in comparison to the normal galaxy population. In addition, we explore the relation between AGN luminosity and molecular gas content \citep[e.g.][]{Maiolino:2007} which is less affected by the timescale issue than the measurement of the SFR.

Throughout the paper we assume a concordance cosmological model with $H_0=70\,\mathrm{km}\,\mathrm{s}^{-1}\,\mathrm{Mpc}^{-1}$, $\Omega_{\mathrm{m}}=0.3$, and $\Omega_\Lambda=0.7$.

\section{The low-redshift QSO sample}\label{sect:sample}
Our low-redshift QSO sample ($z<0.2$) is drawn from the Hamburg/ESO survey \citep[HES,][]{Wisotzki:2000}  based on an early QSO catalog of \citet{Koehler:1997} covering an area of $611\,\mathrm{deg}^2$ on the sky. It is a statistically complete flux-limited sample consisting of the 18 brightest QSOs taking into account the variable depth of individual HES survey fields. The QSOs therefore correspond to the most luminous QSOs in the respective redshift range. Optical to near-IR multi-band imaging data of the sample was presented and discussed in \citet{Jahnke:2004b}. Those data provide morphological information of the host galaxies and host galaxy colors, from which stellar masses were estimated through spectral energy distribution modeling. The effective radii of those massive QSO host galaxies varies between $1\,\mathrm{kpc}<R_\mathrm{e}<10\,\mathrm{kpc}$ corresponding to angular sizes of $4''$ up to $20''$ on sky.

Follow-up optical IFU spectroscopy for 18 objects of the QSO sample was recently presented in Paper~I. Those data already added crucial information about the presence, extent and physical conditions of the ionized gas in the host galaxies (e.g. its ionization state and metallicity). We refer to Paper~I for additional information on the basic sample properties. In this paper, we focus on the characteristics of ongoing star formation, based on the H$\alpha$ surface brightness and kinematics inferred from the IFU data, as described in the following section. We further complement the multi-colour imaging and IFU data of the sample with single-dish \COzero\ and \COtwo\ observations that probes the molecular gas content of the QSO hosts.

\subsection{Optical IFU data and the H$\alpha$ kinematics}
The optical IFU observations were carried out with VIsible MultiObject Spectrograph \citep[VIMOS,][]{LeFevre:2003} at the Very Large Telescope during period 72 and 83. The VIMOS Field-of-View (FoV) was set to $27''\times27''$ or $13''\times13''$ with a spatial sampling of $0.66''$ or $0.33''$, respectively, in order to well cover the entire QSO host galaxies. We chose one of the HR blue, HR orange, or HR red grisms or a combination of two spectral setups to cover at least the wavelength range from H$\beta$ to [NII]\,$6584$\AA\ at a spectral resolution of $3$\AA\ (FWHM). Details on the observations and data reduction of the VIMOS IFU data can be found in Paper~I.

\begin{table*}
\caption{Details of the IRAM 30\,m observations and ambient conditions. For each object we list the Hamburg/ESO survey name and the morphological classification in brackets as B for bulge-dominated, D for disc-dominated and M for major merger. We also list the dates of the observing night(s), range in telescope elevations, on-source integration time ($t_\mathrm{exp}$) per night, central frequencies for the E090 and E230 bands, the range in precipitable water vapour during observation, estimated rms in the E090 band for a channel width of 50km/s for the combined integration time, and finally some remarks on the general observing conditions.}
\label{tab:obs_log}
\input{obs_log.tex}

\end{table*}

As described in Paper~I, it is crucial to disentangle the bright emission of the type-1 QSO nucleus from the host galaxy in the optical IFU data. We used our iterative algorithm \QDeb\ for this task which we introduced in \citet{Husemann:2013a} and applied to the VIMOS observation of this sample as described in Paper I. The algorithm relies on the fact the the broad Balmer lines originate from the AGN broad-line region emitted on $<$\,1\,pc scales \citep[e.g.][]{Peterson:2004,Kaspi:2005} even for the most luminous QSOs. Hence, the Point-Spread Function (PSF) can be directly re-construct for a given observation at the observed wavelength of the broad emission line \citep[e.g.][]{Jahnke:2004}. Based on the re-constructed PSF we iteratively subtracted the beam-smeared QSO spectrum from the IFU data taken into account the host galaxy contamination to the QSO spectrum itself. As a result we obtained a QSO and a host galaxy datacube. We modeled and subtracted the stellar continuum from the host galaxy datacube with the \textsc{starlight} spectral synthesis code \citep{CidFernandes:2005,CidFernandes:2013a} and then co-added the spectra of independent emission-line regions across the host galaxies. Based on the [O\,III]\,$\lambda$5007/H$\beta$ vs. [NII]\,$\lambda6583$/H$\alpha$ line ratio diagnostic diagram \citep{Baldwin:1981, Veilleux:1987} we distinguished between H\,II region complexes powered by ongoing star formation and extended AGN-ionized regions on kpc scales. Based on this analysis we reported star formation rates (SFRs) calculated from the extinction-corrected H$\alpha$ luminosity of the H\,II regions in Paper~I using the conversion by \citet{Kennicutt:1998} and provide upper limits for host galaxies dominated by AGN photoionization throughout their galaxy.

The SFR and the specific SFR (sSFR), which is the SFR divided by the stellar mass, are the prime quantities from the IFU data that we take from Paper~I. In addition,  we construct and present the spatially resolved H$\alpha$ velocity field of all host galaxies in this paper. We model the H$\alpha$ line and the [NII]\,$\lambda\lambda6548,6583$ doublet as a system of single Gaussian emission lines with common velocity and intrinsic velocity dispersion as well as a fixed [NII]\,$\lambda6548$/[NII]\,$\lambda6583$ ratio of 1/3. Errors on all parameters are estimated using a Monte Carlo approach. A sample of 50 datacubes were created for which each pixel value was randomly varied within its error computed from the photon, read-out and background noise by the data reduction pipeline. The QSO-host deblending and stellar continuum subtraction was done on each of the cubes independently and we estimate the uncertainty of the emission-line fitting per spaxel  as the standard deviation of results from the 50 Monte-Carlo realizations. We assume that a reliable measurement for a spaxel is achieved at a H$\alpha$ signal-to-noise of 5, a radial velocity error of $<$\,20\,km/s, and a velocity dispersion error of $<$\,50\,km/s. The cleaned maps are shown in Fig.~\ref{fig:overview}. For comparison with the integrated CO line spectrum we also construct the spatially-integrated emission-line H$\alpha$ profile from the spatially-resolved kinematics. The resulting line profile is also shown in Fig.~\ref{fig:overview} (red line in left panel), integrating all regions identified to be powered by star formation which sometimes subtended the entire host galaxy. 

A more detailed analysis of the ionized gas kinematics comparing different emission lines like H$\alpha$ and \Ox\ and looking for multiple kinematic components is beyond the scope of this paper. It will be subject of a dedicated paper with the emphasis of non-circular motions as potential signatures of interactions and/or ionized gas outflows. Here, we restrict ourselves entirely on the comparison of the ionized and molecular gas properties.

\subsection{Observations and data reduction of CO line observations}
\begin{figure*}
\includegraphics[width=\textwidth]{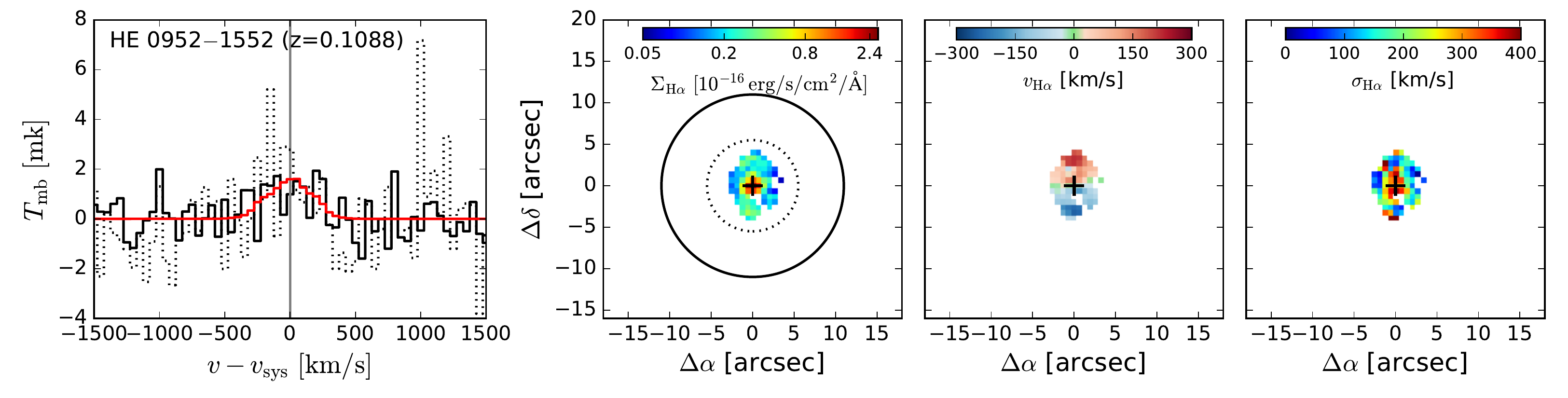}\vspace*{-2mm}
\includegraphics[width=\textwidth]{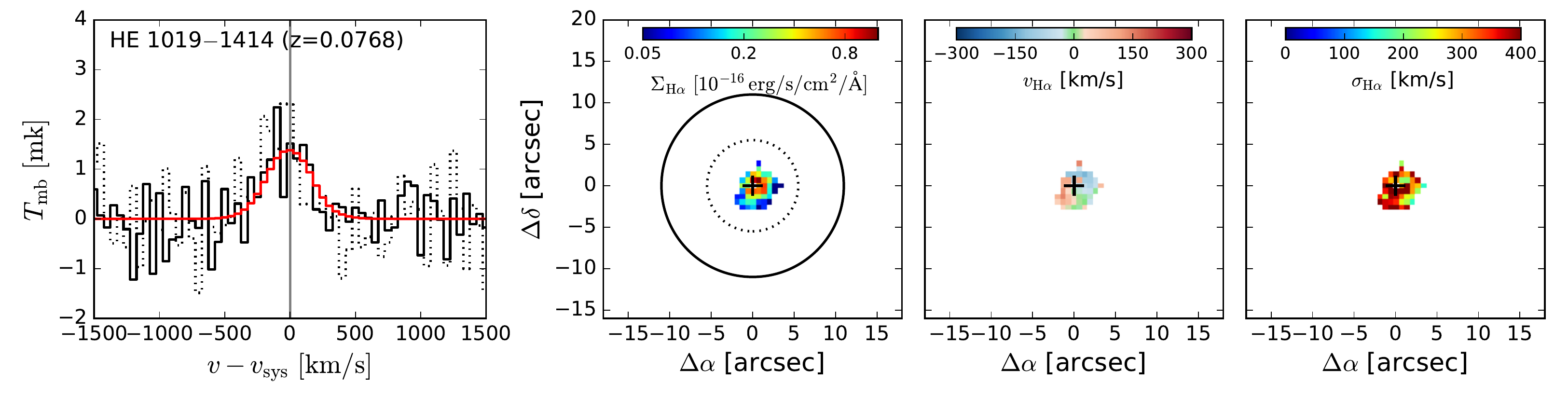}\vspace*{-2mm}
\includegraphics[width=\textwidth]{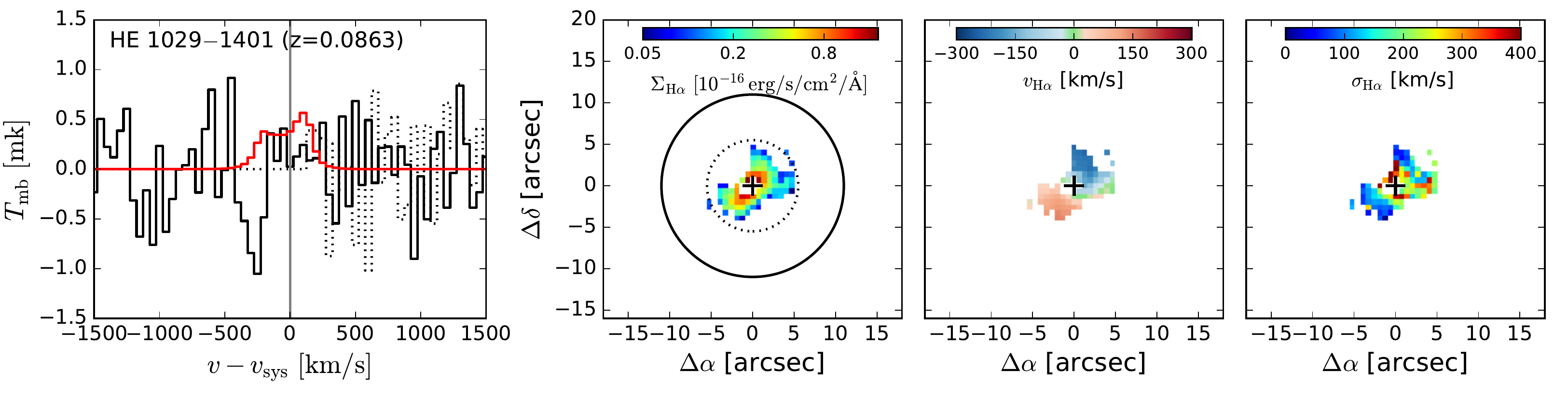}\vspace*{-2mm}
\includegraphics[width=\textwidth]{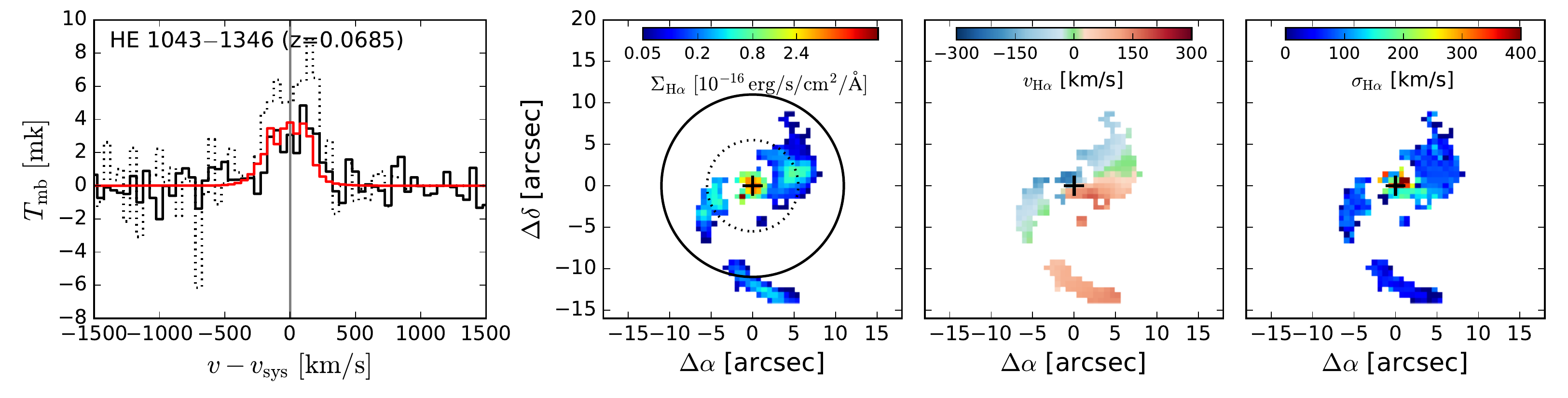}\vspace*{-2mm}
\includegraphics[width=\textwidth]{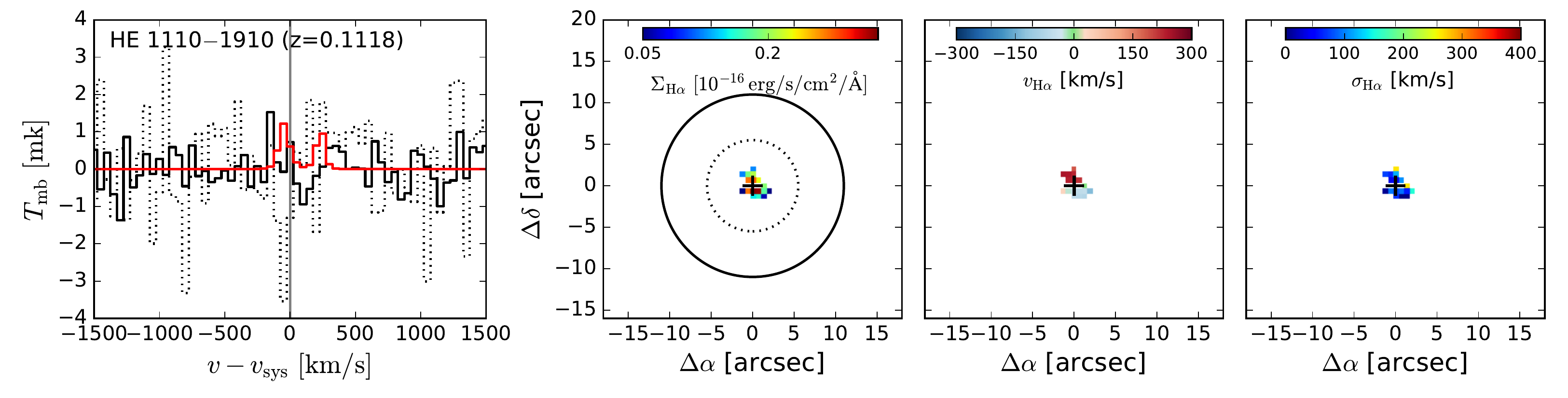}\vspace*{-2mm}

\caption{Comparison of the \COzero\ and \COtwo\ lines (detected or undetected) with the spatially integrated H$\alpha$ line properties derived from the VIMOS IFU data after subtraction of the QSO emission. \textit{Left panels:} \COzero\ line profile (solid black line) and \COtwo\ line profile (dotted black line) with respect to integrated H$\alpha$ line profile (solid red line) matched in equivalent width on the plot. The systemic redshift used here is the median radial velocity in the H$\alpha$ line maps and marks the zero-velocity reference of the line profile highlighted by the vertical gray solid line. \textit{Right panels:} The H$\alpha$ surface brightness distribution, the H$\alpha$ radial velocity with respect to the systemic one, the H$\alpha$ velocity dispersion as measured from the VIMOS IFU are shown from left to right, respectively.}
\label{fig:overview}
\end{figure*}

\begin{figure*}
\includegraphics[width=\textwidth]{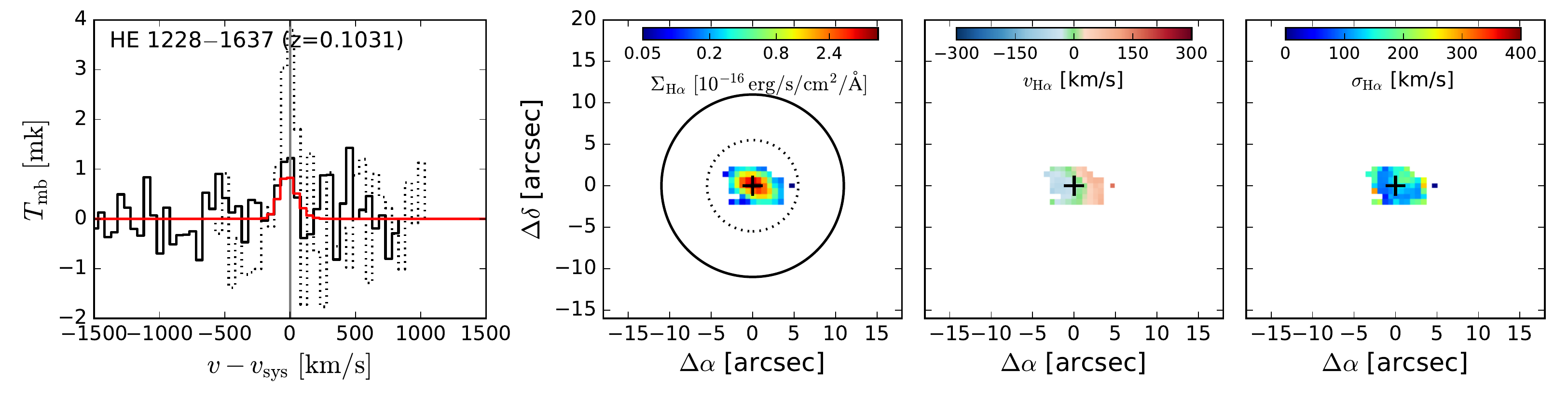}
\includegraphics[width=\textwidth]{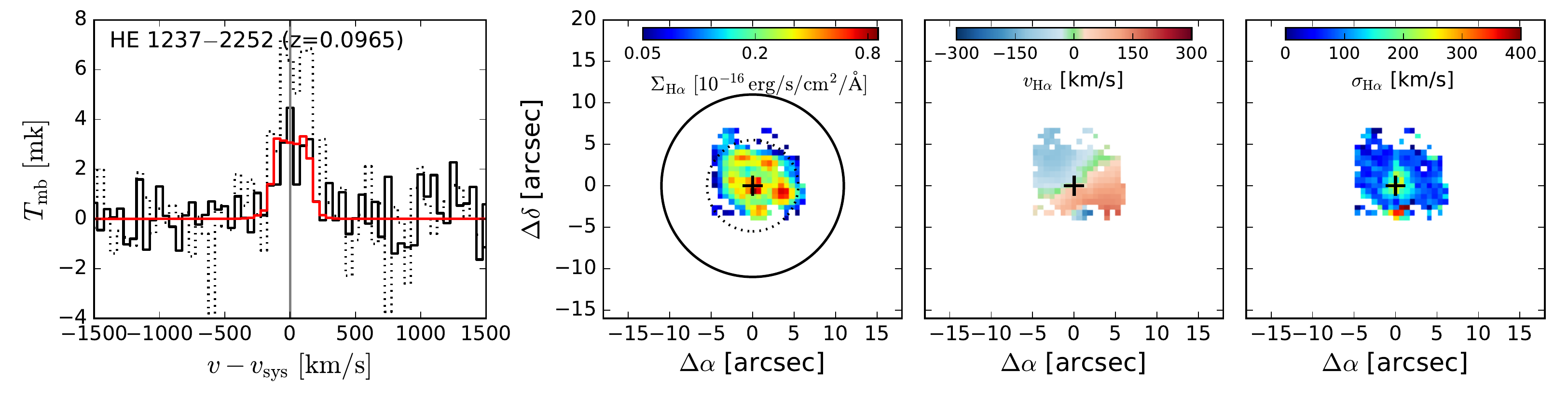}
\includegraphics[width=\textwidth]{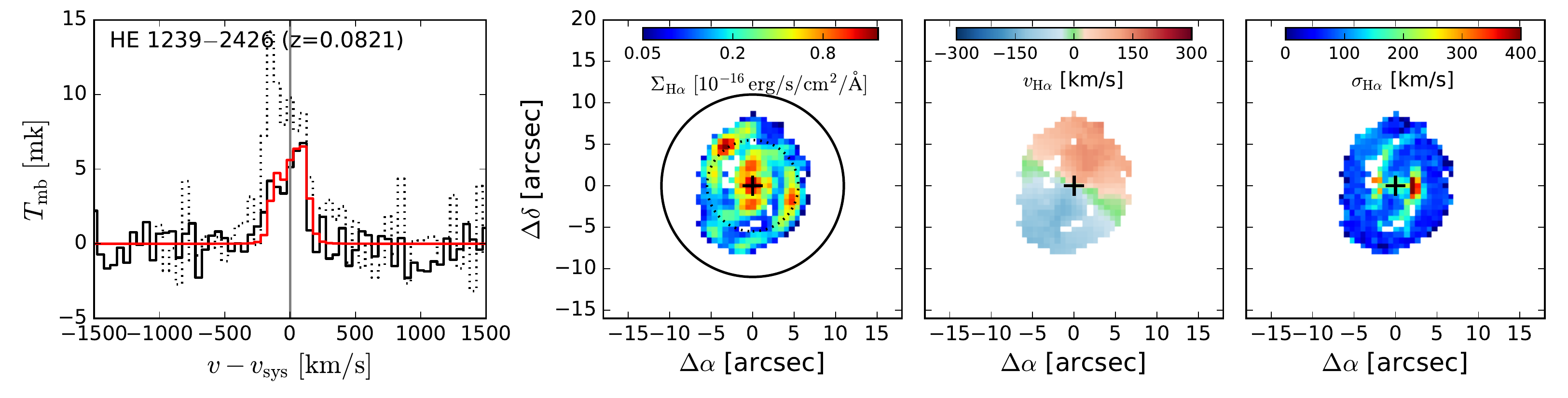}\vspace*{-0.7cm}
\includegraphics[width=\textwidth]{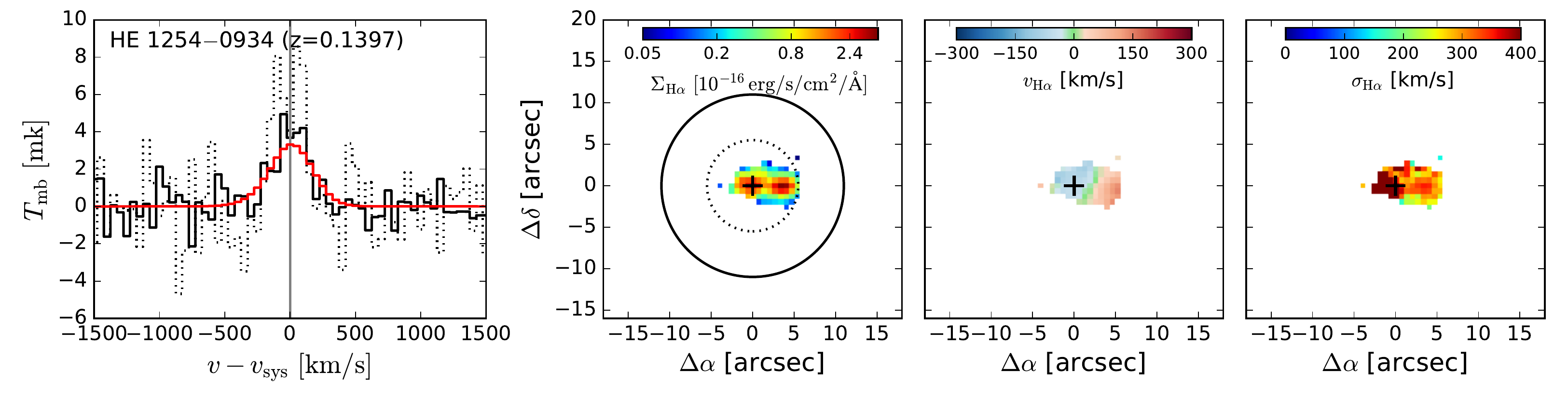}
\includegraphics[width=\textwidth]{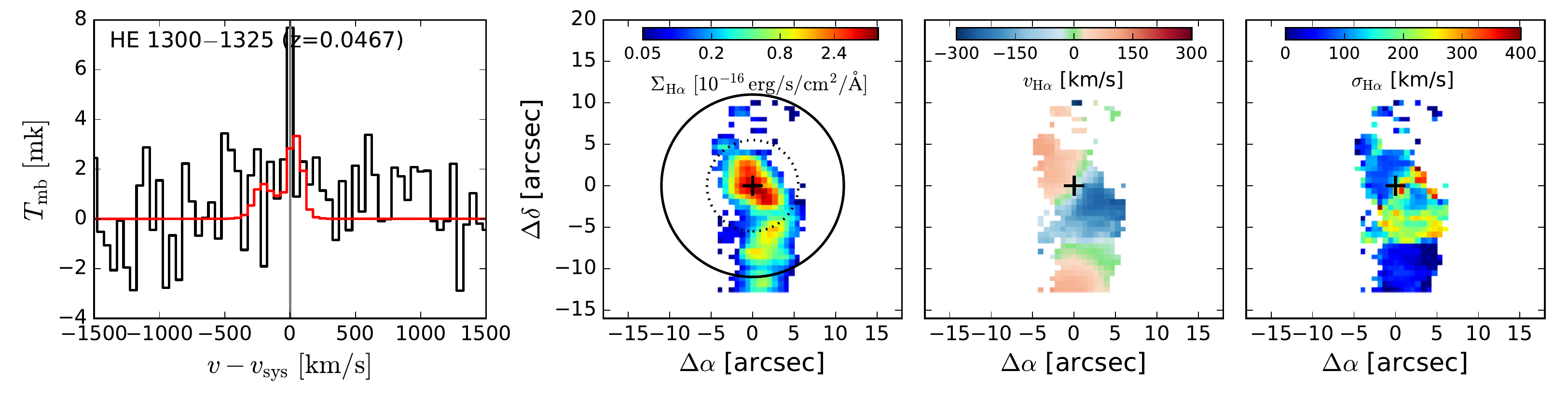} 
\contcaption{}
\end{figure*}

\begin{figure*}
\includegraphics[width=\textwidth]{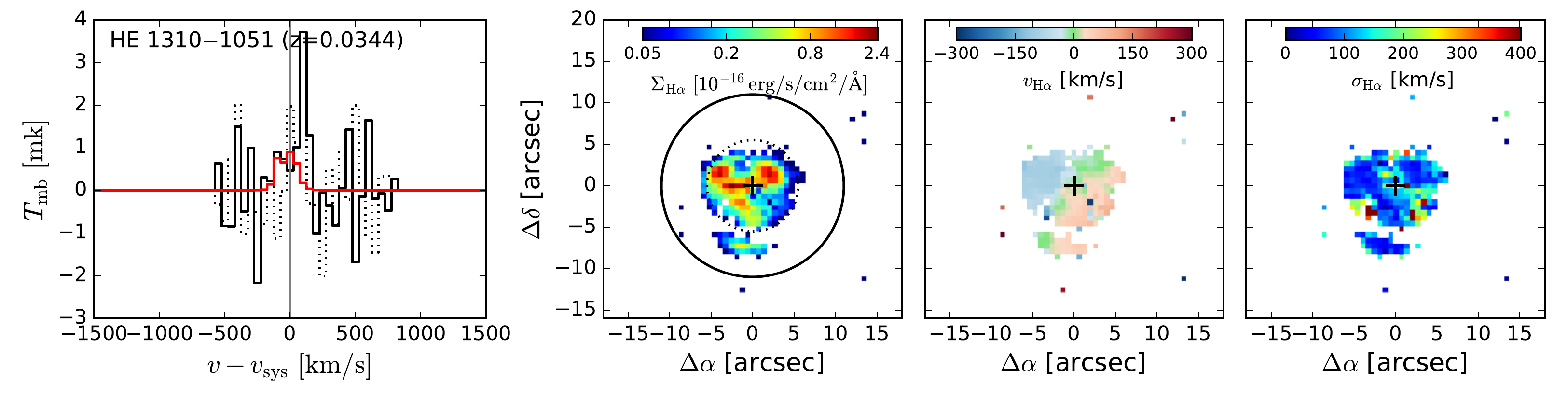}
\includegraphics[width=\textwidth]{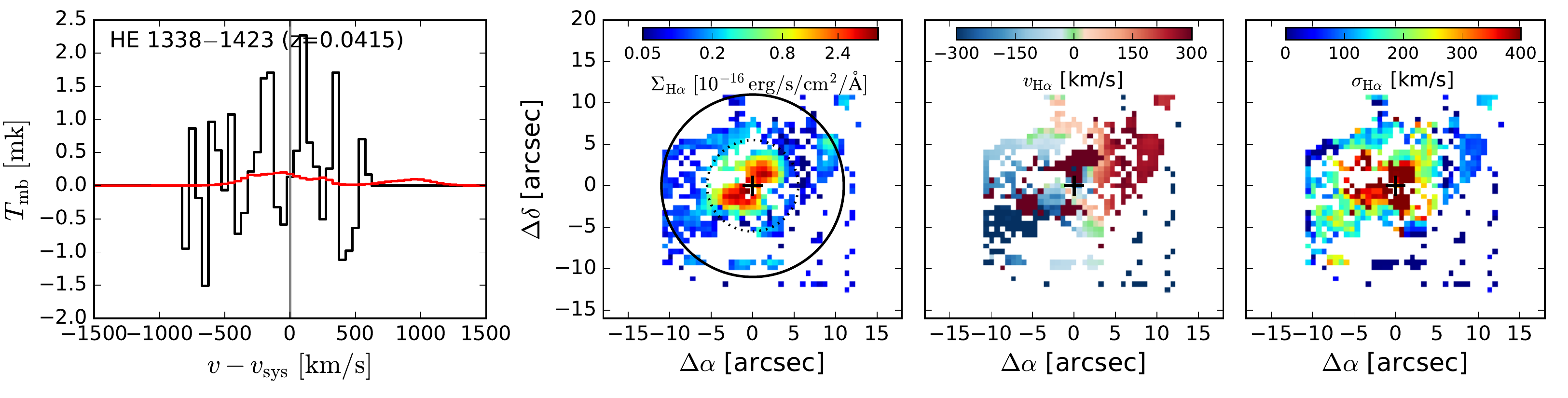}
\includegraphics[width=\textwidth]{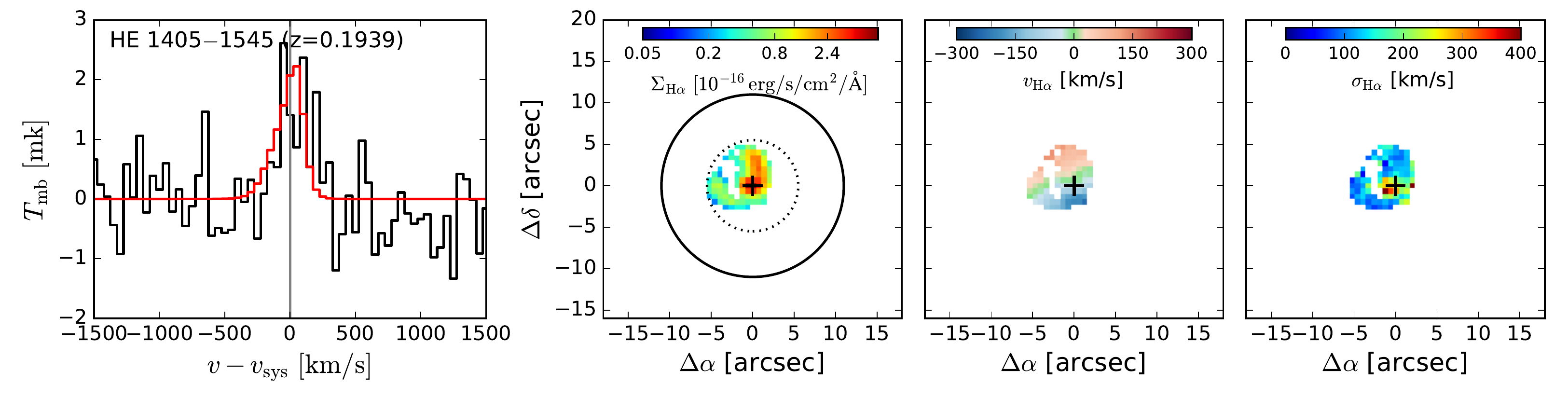}
\includegraphics[width=\textwidth]{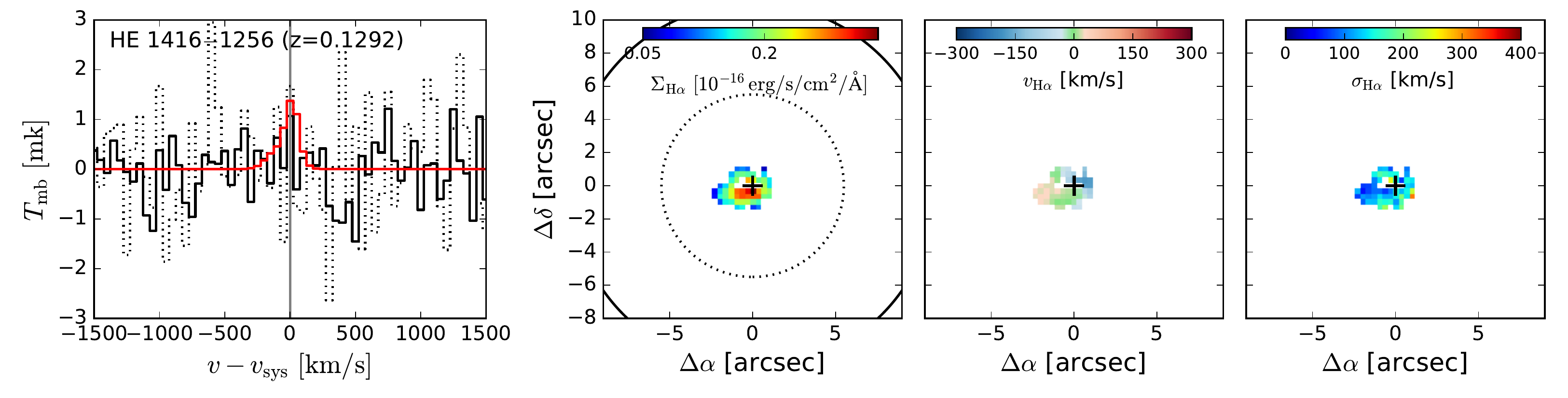}
\contcaption{}
\end{figure*}
Single dish observations for a sub-set of the QSO sample were carried out with the IRAM 30\,m telescope at Pico Veleta (Granada, Spain) in January 2014. We selected 12 QSOs for which we estimated a SFR of $0.5\,\mathrm{M}_\odot\,\mathrm{yr}^{-1}$ or an upper limit above this threshold. In addition, two QSOs, HE~1310$-$1051 and HE~1338$-$1423, had previously been targeted with the IRAM 30\,m telescope \citep{Bertram:2007} using the ABCD backend providing a total sample of 14 QSOs with \COzero\ and VIMOS observations. The IRAM 30m observations were carried out with the Eight MIxer Receiver \citep[EMIR,][]{Carter:2012} in dual-band mode to simultaneously observe the redshifted \COzero\ and \COtwo\ lines in the 3\,mm (E090) and 1\,mm (E230) bands. Given the relatively large redshift range covered by the sample we defined a few single side-band tunings such that both CO lines fall within the large $4\times4$\,GHz bandwidth (200\,kHz sampling) of the Fourier Transform Spectrometer (FTS) for several objects.  This was possible due to the accurate redshift information provided by the optical IFU data and allowed us to reduce the overhead time substantially compared to individual tunings per QSO.  We also recorded data with the WILMA backend as a backup, which has a smaller bandwidth of 4x2GHz in single-sideband mode. Thus, it does not cover the lines for all objects at a given tuning.

Since the apparent size of the QSO host galaxies ($<20''$) are smaller than the telescope half power beam size of $\sim22''$ in the 3\,mm band, we only obtained single pointings on each source. Switching between on and off source was done with the wobble switching secondary mirror running at a throw of $1'$ in azimuth and altitude at a frequency of 1\,Hz. We achieved an on-source integration time per object ranging from 23\,min to 3\,h. The exposure time per galaxy was estimated to achieve a S/N$>$3 in \COzero\ based on the SFR and assuming a gas depletion time of 2\,Gyr \citep{Bigiel:2008}. In total we acquired 15.4\,h of on-source integration time for the entire observing run. The observing conditions varied significantly between the observing sessions ranging from excellent conditions with pwv$<$1\,mm and poor conditions with pwv$>$10mm together with strong winds close to the pointing limit of the telescope. An overview of the observations including tuning frequencies and observing conditions for individual observing blocks are reported in Table~\ref{tab:obs_log}. 

We reduced all the data with the software GILDAS/CLASS\footnote{available at http://www.iram.fr/IRAMFR/GILDAS}. Because the FTS bandwidth is split up into three bandpasses/amplifiers, we subtracted independent linear baselines for the corresponding frequency ranges in each polarization. The combined spectrum was then computed as the variance-weighted average of all high quality scans independent of polarization. Finally, the spectrum were binned in frequency so that the final sampling corresponds to $50\,\mathrm{km}\,\mathrm{s}^{-1}$ in the rest-frame of the object. We assume main beam efficiencies of $B_\mathrm{eff}=0.78$ and forward beam efficiencies of $F_\mathrm{eff}=0.94$ at $\nu\sim90\mathrm{GHz}$ to convert the telescope temperatures to main beam temperatures.  Similarly, we assume  $B_\mathrm{eff}=0.63$ and $F_\mathrm{eff}=0.94$ at $\nu\sim210\mathrm{GHz}$. The resulting spectra covering the \COzero\ and, if possible, the \COtwo\ lines are shown in Fig.~\ref{fig:overview} (left panels). 

We measure the velocity integrated line fluxes by fitting a Gaussian function to the CO line profiles. A few profiles clearly show a horn profile, but we still model them with a Gaussian function for consistency. This approach is valid in our case, because the measurement errors for our spectra are larger than the systematic uncertainties of the line profile mismatch. We fix the redshift of the lines to the systemic redshift based on the H$\alpha$ kinematics, which reduces the number of free parameters and provides a more robust line flux measurement. A Monte Carlo approach is used to estimate the uncertainties on our CO line fluxes. We generate 200 representation of each spectrum by randomly drawing values from a normal distribution centred on the original spectrum with standard deviation set by the noise. The \COzero\ line is detected in 8 out of 14 QSO host galaxies with $\gtrsim3\sigma$ and the \COtwo\ line is detected in 7 of 9 cases where the line could be covered by the backends. We estimate $3\sigma$ upper limits for the non-detections assuming a Gaussian profile again with a FWHM of 350\,km/s which is the mean value from all our detections. In Table~\ref{tab:line_measurements}, we report the results of those line measurements.

\begin{table*}
\caption{Results of the ${}^{12}\mathrm{CO(1-0)}$ and ${}^{12}\mathrm{CO(2-1)}$ line measurements. All upper limits are provided as $3\sigma$ detection limits. }
\label{tab:line_measurements}
\input{sample_results.tex}
\begin{flushleft}
$^{*}$ \COzero\ line width fixed to the measured width of \COtwo\ given its robust detection which provides an accurate prior.
\end{flushleft}
\end{table*}

\section{Results}\label{sect:results}
\subsection{Molecular gas mass and gas fractions}
Here we follow the prescription of \citet{Bertram:2007} to estimate the cold molecular gas mass $M(\mathrm{H}_2)$ from the \COzero\ line as described in \citet{Solomon:1992a}. First, we convert the \COzero\ velocity integrated line flux to a line luminosity,\begin{equation}
 L^\prime_\mathrm{CO}(1-0) = 23.5\Omega_\mathrm{S*B}D_\mathrm{L}^2 I_\mathrm{CO}(1-0)(1+z)^3\quad ,
\end{equation}
where $\Omega_\mathrm{S*B}$ is the solid angle of the source convolved by the solid angle of the beam in square arcsecond, $D_\mathrm{L}$ is the luminosity distance and $z$ is the systematic redshift of the object. Because the solid angle of all source, $\Omega_\mathrm{S}$, is significantly  smaller than the beam size $\Omega_\mathrm{B}$ at least for the \COzero\ line, we adopt the simplification to assume that $\Omega_{\mathrm{S*B}}\approx\Omega_\mathrm{B}$. The \COzero\ line at the redshift of our targets fall in the 3mm band at which the IRAM telescope has a half-power beam size of $\sim22''$.

The total molecular hydrogen gas mass is calculated by multiplying $L^\prime_\mathrm{CO}$ with an appropriate scale factor $\alpha_\mathrm{CO}$ \citep[see][and references therein]{Bolatto:2013}. This conversion factor has been empirically determined from individual molecular gas clouds in the Milky Way and nearby galaxies $\alpha_\mathrm{CO}\sim3.2\,M_{\sun}\,\mathrm{K}^{-1}\,\mathrm{km}^{-1}\,\mathrm{pc}^2$ without correction for Helium \citep[e.g.][]{Dickman:1986,Strong:1988,Strong:1996,Blitz:2007}. On the other hand, a significant dependence of $\alpha_\mathrm{CO}$ has been reported with metallicity \citep[e.g.][]{Boselli:2002,Leroy:2011} and with starburst environments, e. g. ultra-luminous infrared galaxies (ULRIGs), for which a significantly lower conversion factor has been obtained $\alpha_\mathrm{CO}\sim0.8\,M_{\sun}\,\mathrm{K}^{-1}\,\mathrm{km}^{-1}\,\mathrm{pc}^2$ \citep[e.g.][]{Solomon:1997}. Since luminous QSOs had been thought to be usually associated with gas-rich mergers of (U)LIRG type most studies on the molecular gas content of QSO host galaxies have used a low $\alpha_\mathrm{CO}\sim0.8\,M_{\sun}\,\mathrm{K}^{-1}\,\mathrm{km}^{-1}\,\mathrm{pc}^2$ \citep[e.g.][]{Riechers:2006,Villar-Martin:2013} consistent with starburst systems. 

This merger-driven scenario has been challenged in recent years since a large fraction of QSOs appear not be hosted by major mergers, but undisturbed disc galaxies instead \citep[e.g.][]{Cisternas:2011,Schawinski:2012,Villforth:2014,Mechtley:2016}. Indeed, most of the QSO hosts in our sample appear as undisturbed disc or elliptical galaxies, so that it is reasonable to assume a Galactic scale factor of $\alpha_\mathrm{CO}=4.35\,\mathrm{M}_\odot\/(\mathrm{K}\,\mathrm{km}\,\mathrm{s}^{-1}\,\mathrm{pc}^{2})$ that also takes into account the Helium abundance. Such a conversion factor has also been used by studies of the molecular gas in low-redshifts QSOs \citep[e.g.][]{Evans:2001,Scoville:2003}, in the local galaxy population by the COLD~GASS survey \citep{Saintonge:2012} and in massive elliptical galaxies by the ATLAS${}^\mathrm{3D}$ survey \citep{Young:2011}. We therefore (re)compute all molecular gas mass with this conversion factor to be consistent throughout the article.

In order to compare the properties of our QSO hosts with non-AGN galaxies we use the COLD~GASS survey and ATLAS${}^\mathrm{3D}$ survey as comparison samples for disc and bulge-dominated galaxies, respectively. The COLD~GASS survey is an IRAM Legacy survey that observed a representative sample of $\sim$350 nearby galaxies with stellar masses $>10^{10}M_\odot$ in the \COzero\ line. It is one of the largest and most homogeneous databases for the molecular gas content of nearby galaxies to date. It has a \COzero\ 5$\sigma$ detection limit corresponding to about 1-10\% of the corresponding stellar masses. We cross-match the COLD~GASS sample with the visual classification of galaxies from the Galaxy Zoo 2 project \citep{Willett:2013} to distinguish also between disc- and bulge-dominated systems (S and E morphological classes), which leads to a sub-sample of 274 galaxies. Bulge-dominated galaxies have often lower molecular gas fractions than the COLD~GASS detection limit. This gap is filled by the ATLAS${}^\mathrm{3D}$ survey, which is an IFU survey of a volume limited sample of nearby bulge-dominated galaxies complemented by deep \COzero\ observation down to gas fractions of 0.1\% \citep{Young:2011}. 

\begin{figure}
\resizebox{\hsize}{!}{\includegraphics{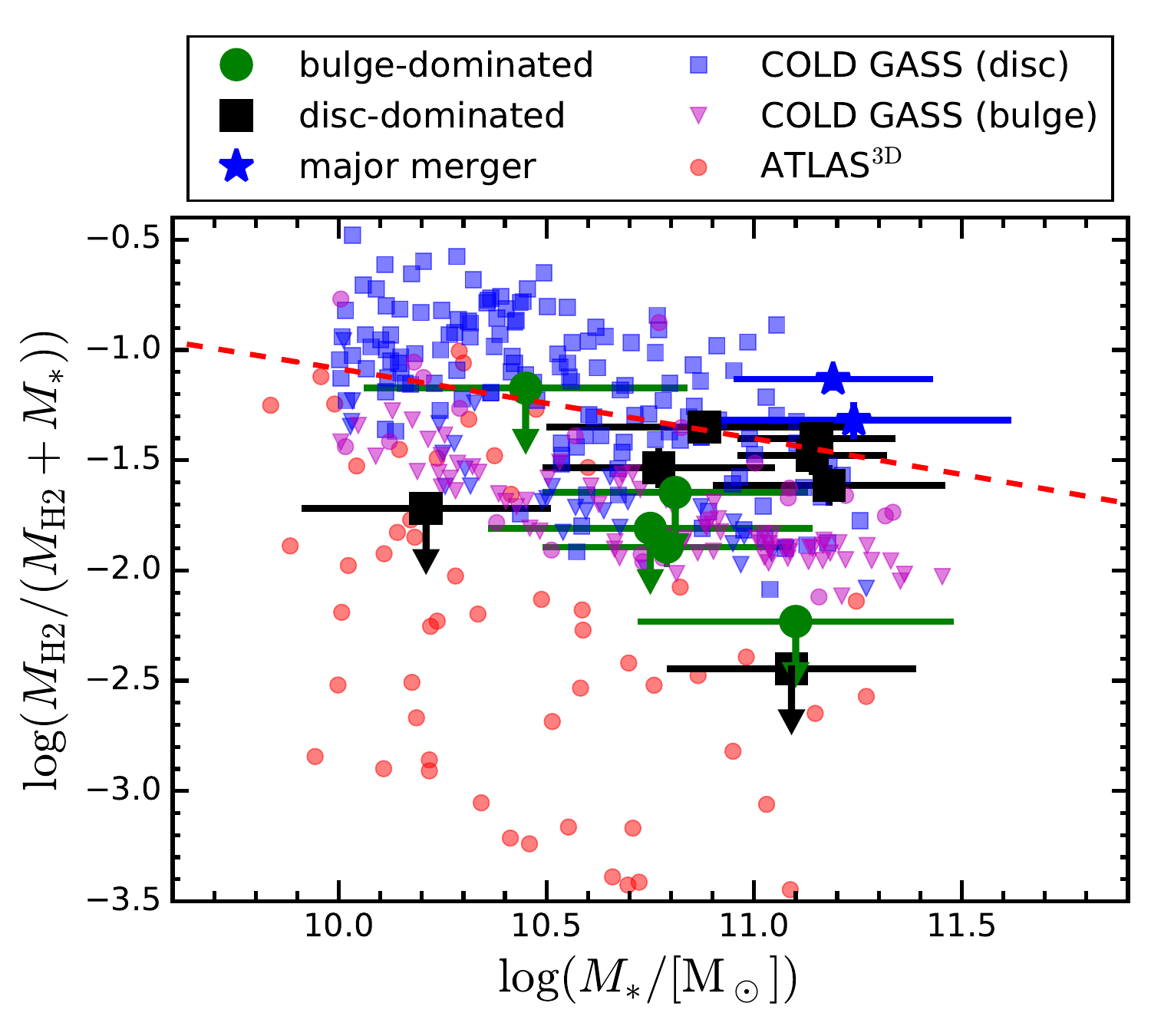}}
\caption{Comparison of molecular gas fraction against stellar mass for our luminous QSOs and non-AGN galaxies. The QSO host galaxies are divided into disc-dominated, bulge-dominated and ongoing major merger systems. Comparison samples of non-AGN galaxies are drawn from the COLD~GASS survey, and ATLAS${}^\mathrm{3D}$ for bulge-dominated galaxies. The dashed line is a parametrization of the gas fraction for the galaxy population as derived by \citet{Popping:2012} for redshift $z=0.06$.}
\label{fig:gas_fraction}
\end{figure}

The majority of our QSO host galaxies contain substantial amounts of molecular gas ranging between $10^9$--$10^{10}\,\mathrm{M}_\odot$. This is in agreement with previous studies of the molecular gas content in luminous low-redshift QSOs \citep{Scoville:2003,Bertram:2007,Evans:2006,Villar-Martin:2013}. We also provide stringent upper limits for the non-detection corresponding to $\lesssim10^9\,\mathrm{M}_\odot$ ($3\sigma$). Most of the upper limits originate from bulge-dominated QSO host galaxies which therefore contain systematically less molecular gas compared to the disc-dominated hosts and ongoing major mergers.  

We present molecular gas fractions ($M_{\mathrm{H}2}/(M_{\mathrm{H}2}+M_*)$) in Fig.~\ref{fig:gas_fraction} against the total stellar mass ($M_*$) for our sample compared to the overall non-AGN  galaxy population as seen by COLD~GASS and nearby bulge-dominated systems as observed by ATLAS${}^\mathrm{3D}$. To statistically compare the distributions including the censored data, we take out the mass-dependence of the gas fraction using the parametrization derived by \citet{Popping:2012},
\begin{equation}
 f_\mathrm{gas,Pop12}=\frac{M_{\mathrm{H2}}}{M_{\mathrm{H2}}+M_*} = \frac{1}{\exp^{(log M_*-A)/B}+1}\quad,
\end{equation}
where the best-fitting parameter are found to be $A=6.15(1+z/0.036)^{0.144}$ and $B=1.47(1+z)^{-2.23}$. Based on this mean relation of gas fractions for the overall galaxy population  we define the offset from the relation as $\Delta f_\mathrm{gas} = f_\mathrm{gas,observed} - f_\mathrm{gas,Pop12}$.

Here, we use the \textit{survival package} of \textsc{R} \citep{R:2013} and apply the logrank test and also the modified Gehan Wilcoxon test to infer whether two samples are drawn from the same parent sample taking into account censored data. We find that $\Delta f_\mathrm{gas}$ for the disc-dominated QSO host galaxies is statistically consistent with the distribution of non-AGN disc galaxies. The bulge-dominated QSO galaxies are also statistically consistent with non-AGN bulge-dominated galaxies from COLD~GASS and ATLAS${}^\mathrm{3D}$. However, their gas fractions are significantly lower than the disc-dominated galaxies at 99\% confidence. The two ongoing major mergers, HE~1254$-$0934 and HE~1405$-$1545, show clearly enhanced molecular gas fractions. To check whether this is caused by a lower $\alpha_\mathrm{CO}$ factor we estimate the dynamical mass of those systems. For HE~1245$-$0934 we measure a radial velocity of $V_r=$120\,km/s at $R_0=12$\,kpc and for HE~1405$-$1545 we obtain $V_r=125$\,km/s at $R_0=14$\,kpc. This corresponds to dynamical masses of $M_\mathrm{dyn}=1.2\times10^{11}M_{\sun}$ for HE~1254$-$0934 and $M_\mathrm{dyn}=1.0\times10^{11}M_{\sun}$ for HE~1405$-$1545, respectively, adopting an inclination of $i=40^\circ$ in both cases. The gas mass is still much smaller than the dynamical mass in both case.  Hence, we continue to apply the same $\alpha_\mathrm{CO}$ conversion factor for these two mergers as for the other galaxies in the sample to be consistent.

\begin{figure*}
 \includegraphics[width=0.49\textwidth]{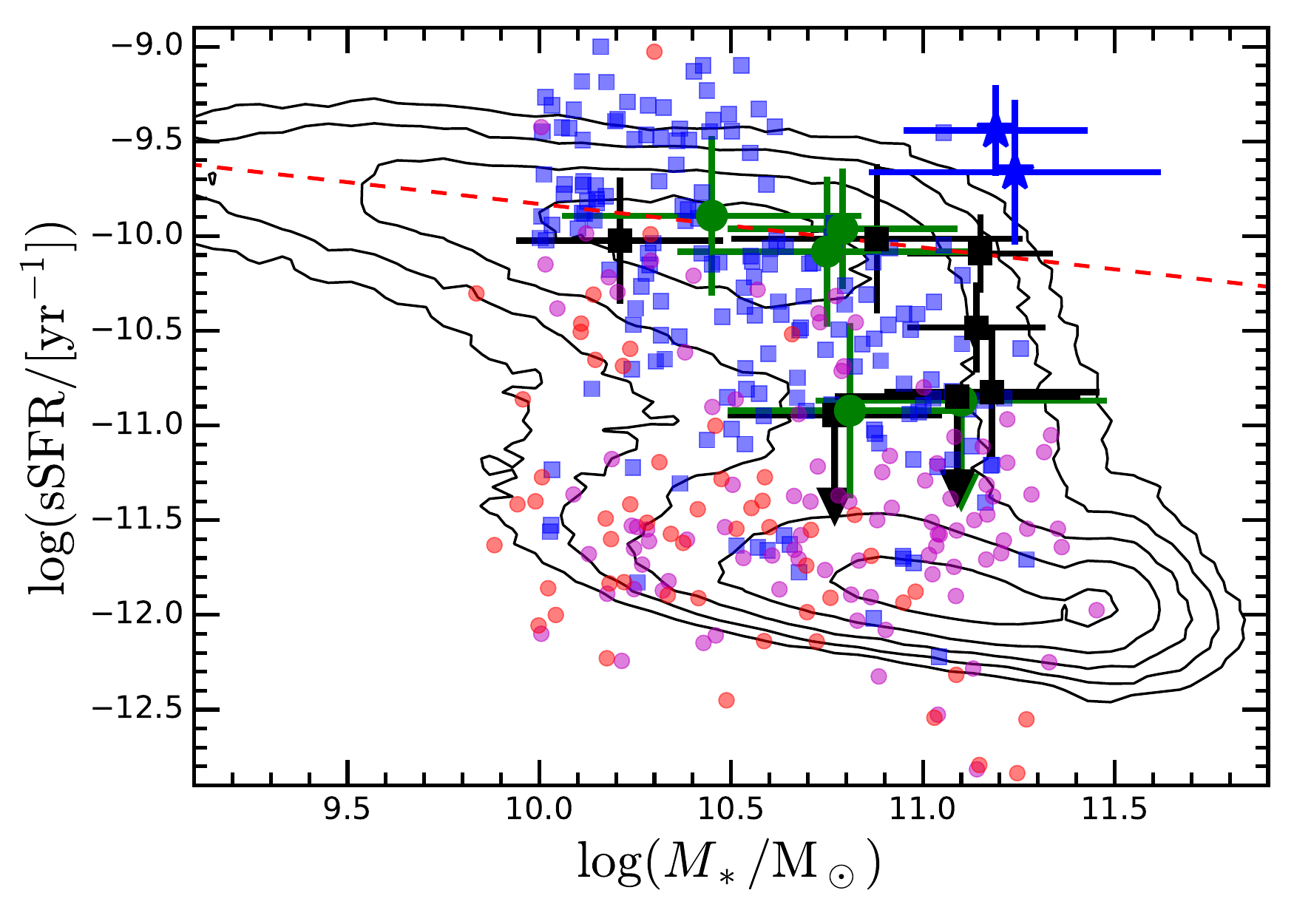}\includegraphics[width=0.49\textwidth]{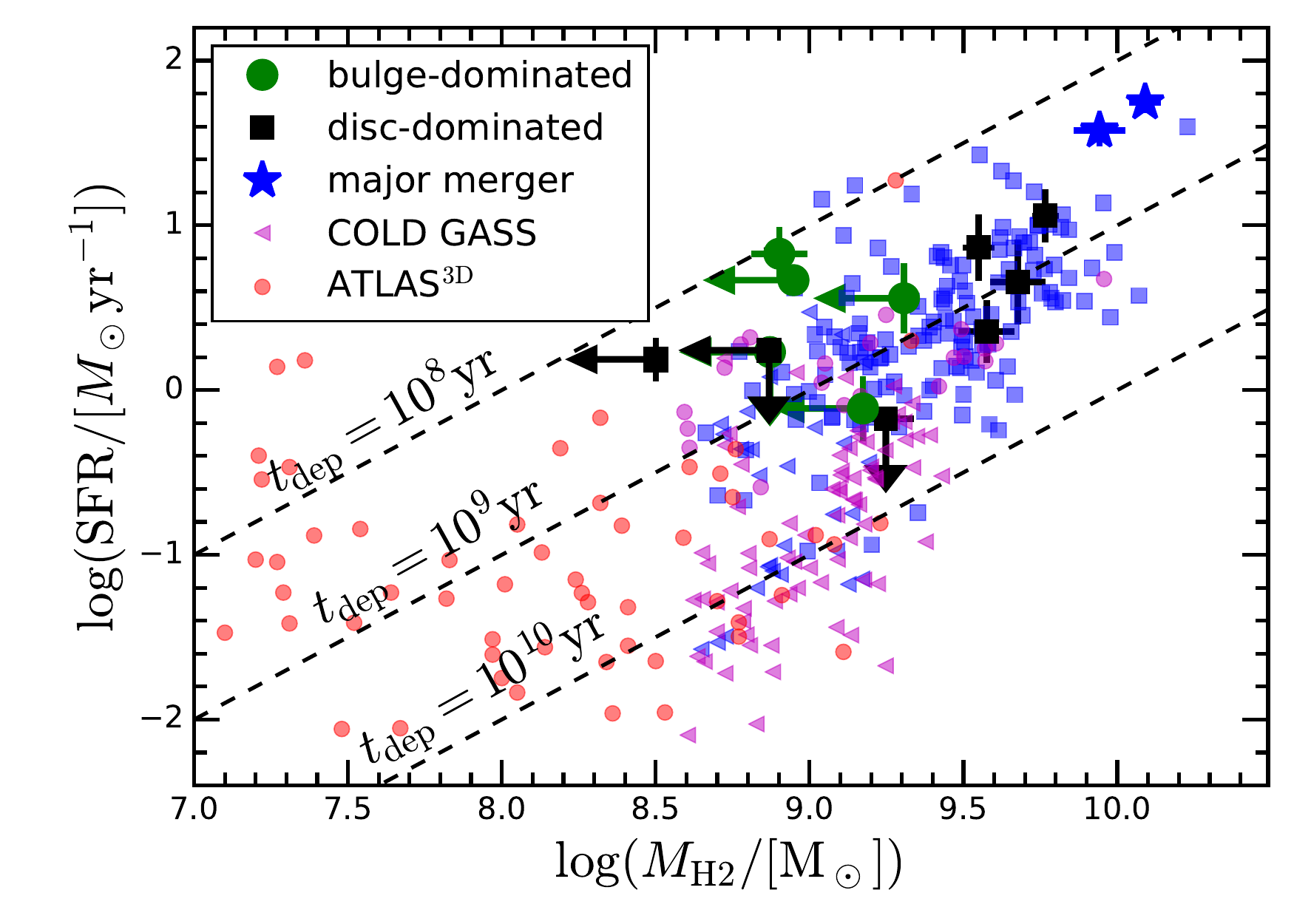}
 \caption{\textit{Left panel:} Specific SFR against stellar mass for our QSO sample. We separated our QSO host galaxies by morphological type as labeled in the right panel. This is similar diagram as presented in Paper I, but restricted to QSO subsample subject of this paper. The underlying contours indicate the distribution of SDSS emission-line galaxies for visual comparison. Most of our QSO host galaxies populate the star forming sequence of galaxies independent of morphological type, but with a few cases of significantly low sSFR. \textit{Right panel:}  SFR against molecular gas content for our QSO sample. The dashed lines indicate the directions of constant molecular gas depletion timescale of $10^8$, $10^9$, and $10^{10}$\,yr as labeled. Most of the disc-dominated QSOs exhibit depletion time scales around $10^9$\,yr typical for normal star forming galaxies. The bulge-dominated and major merger QSO hosts show a tendency towards shorter depletion time scales.}
 \label{fig:SFR_eff}
 \end{figure*}

\subsection{Comparison of the ionized and molecular gas kinematics}\label{sect:kinematics}
A unique advantage of our data set is the possibility to compare the kinematics of the molecular gas traced by \COzero\ with the ionized gas traced by H$\alpha$. For all our galaxies with a \COzero\ detection we compare the \COzero\ line profile with the spatially integrated H$\alpha$ line profile after subtracting the QSO contribution (Fig.~\ref{fig:overview}). In all cases we quantify that the line profiles of \COzero\ and \Ha\ are statistically consistent with each other at 95\% confidence, despite a global velocity offset of about 50\,km/s is seen for HE~1237$-$2252, by computing the reduced $\chi^2$ for the \COzero\ line profile assuming that  H$\alpha$ is a model to the data $\chi_\nu^{\mathrm{H}\alpha}$ (Table~\ref{tab:line_measurements}). The resulting values range between $0.7<\chi_\nu^{\mathrm{H}\alpha}<1.3$ in all cases with detected \COzero\ emission confirming that both line profiles are consistent with each other at the given S/N. 

This comparison highlights that, even without resolving the \COzero\ velocity field, the molecular gas is tracing the spatially-resolved kinematics of H$\alpha$. It implies that the molecular gas is spatially distributed on kpc scales across the galaxy similar to the ionized gas. While this is surprising at first glance given that the molecular gas traces very cold gas with a temperature of a few tens of K and the ionized gas has a temperature of 10\,000\,K, it is actually consistent with the picture that the molecular gas is the seed of ongoing star formation activity. Those very young star forming regions contain massive hot stars that lead to prominent H\,II regions probed through the H$\alpha$ emission. Any necessary difference in the distribution and kinematics of both phases on the scales of molecular clouds ($<$50pc) are basically averaged out due to the kpc resolution of our observations.

In addition, there seems to be no significant additional contribution from circumnuclear molecular gas to the entire budget on sub-kpc scales which we would not be able to resolve in H$\alpha$ due to the subtraction of the point-like QSO spectrum. \textit{We therefore can estimate the gas surface density by adopting the apparent area from the size of the H$\alpha$ emitting region.} This is an essential finding that enables us to study the conditions for star formation by combining both data sets.

\subsection{The star formation efficiency in QSO host galaxies}

In Fig.~\ref{fig:SFR_eff} (left panel), we present the specific SFR against the stellar mass for our QSO sub-sample drawn from Paper I. A similar plot has already been shown in Paper I, but the pre-selection of targets for \COzero\ follow-up based on the expected luminosity introduces a cut at $\log\mathrm{sSFR}\gtrsim-11.0$. Although we exclude four QSO host galaxies due to the expected CO non-detections we still include several targets in-between the blue cloud and passive systems that turned out to be undetected in CO. As already reported in Paper I the majority of our QSO host galaxies lie on the star forming main sequence independent of their host galaxy morphology. To statistically compare the QSO sample with the COLD~GASS and ATLAS${}^\mathrm{3D}$ control samples we determine the difference in sSFR ($\Delta$ sSFR) with respect to the expected sSFR from the local star forming main sequence parametrization of \citet{Elbaz:2007},
\begin{equation}
 \log\mathrm{sSFR} = 0.94+0.77(\log M_*-11.0)-\log M_*
\end{equation}
Applying the logrank  and the modified Gehan Wilcoxon test using \textit{survival package} of \textsc{R} we find that the disc-dominated QSOs are indistinguishable from the star forming disc galaxy population whereas the bulge-dominated QSO exhibit an enhanced sSFR compared to the bulge-dominated non-AGN population at $>$95\% confidence. This is in contrast with the gas fraction of bulge-dominated QSOs being consistent with the comparison sample of the same morphological type. The ongoing major merger systems are clearly above the main sequence as often observed for ongoing gas-rich major mergers \citep[e.g][]{Ellison:2008,Heiderman:2009,Cao:2016}. 

By combining the \COzero\ measurements with our existing IFU observations for this sample, we can compare the \emph{current} SFR against the inferred molecular gas mass (right panel Fig.~\ref{fig:SFR_eff}). The ratio of the two quantities is the gas depletion time $t_\mathrm{dep}=M_\mathrm{H2}/\mathrm{SFR}$. Compared to the control samples from COLD~GASS and ATLAS${}^\mathrm{3D}$ it seems that the majority of our disc-dominated QSO hosts are consistent with a depletion time of $\sim10^9$\,yr. It is very close to normal star forming galaxies probed by COLD~GASS \citep{Saintonge:2011} and measured for various nearby galaxy studies \citep[e.g.][]{Leroy:2008,Bigiel:2008}. Indeed, the logrank test and the modified Gehan Wilcoxon test clearly show that the disc-dominated QSOs and control sample of disc galaxies are statistically indistinguishable in $t_\mathrm{dep}$. The bulge-dominated QSOs appear to have a shorter depletion time compared to the non-AGN bulge-dominated galaxies at 99\% confidence (and also when compared to the non-AGN disc-dominated galaxies at 95\% confidence). This is a direct consequence of the enhanced star formation in the bulge-dominated QSOs while the gas fractions are more similar to non-AGN bulge-dominated galaxies. Our two ongoing major mergers have a shorter depletion time close to $10^8$\,yr which would even be shorter if we assume an $\alpha_{CO}\sim1\,\mathrm{M}_{\odot}/(\mathrm{K}\,\mathrm{km}\,\mathrm{s}^{-1}\,\mathrm{pc}^{2})$ which reduces the molecular gas mass by 0.6\,dex. Such short depletion time scales are commonly observed for gas rich major mergers \citep[e.g.,][]{Gao:2004}.

\begin{figure}
\resizebox{\hsize}{!}{\includegraphics{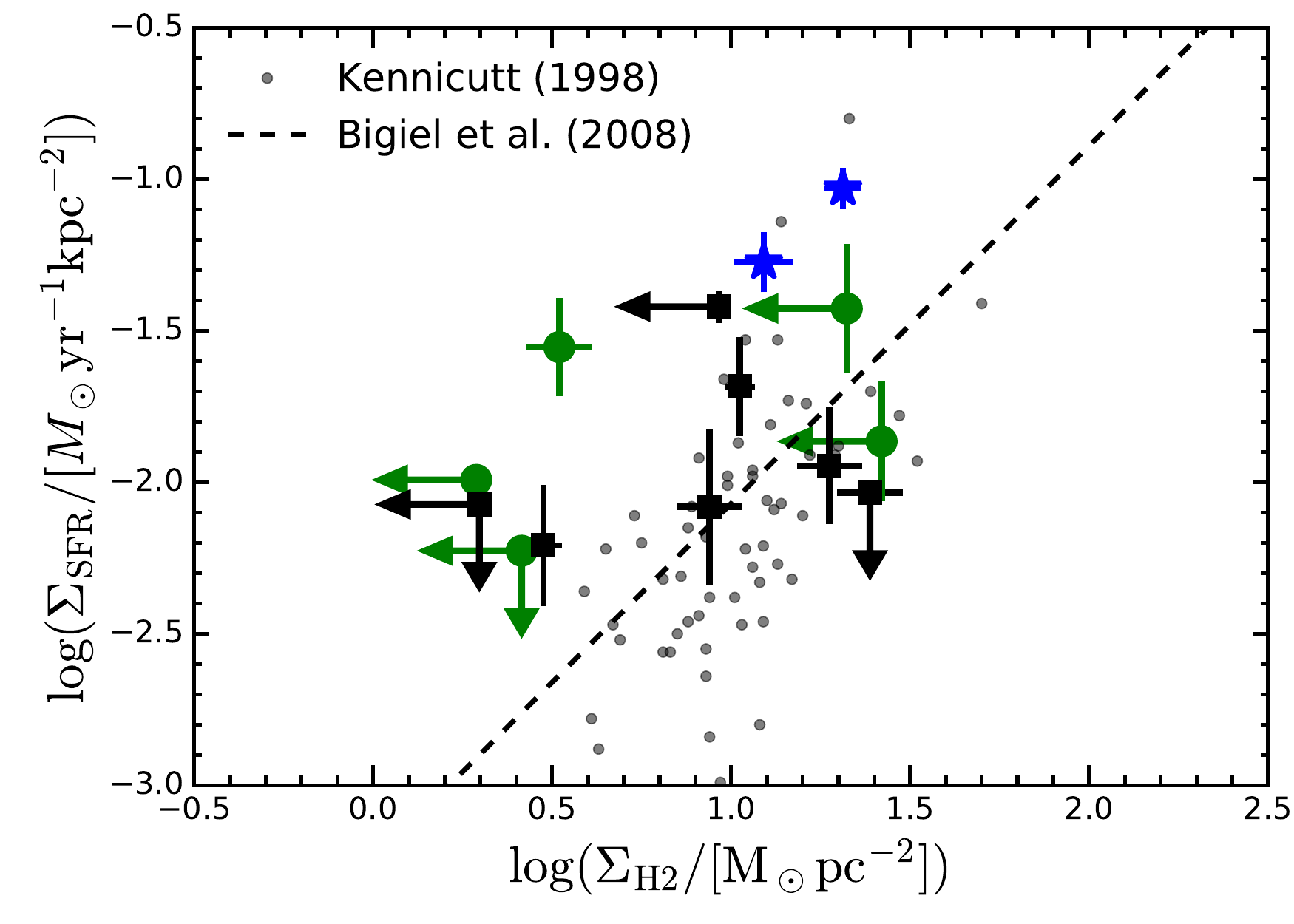}}
\caption{Total surface SFR against surface molecular gas mass for our QSO sample. The symbols denote different morphological types of QSO host galaxies as shown in Fig.~\ref{fig:SFR_eff}. The solid gray circles and dashed line are taken from \citet{Kennicutt:1998} and \citet{Bigiel:2008}, respectively.}
\label{fig:Schmit_law}
\end{figure}

The high SFR of the bulge-dominated QSOs may be caused by a higher gas surface density on average. To test this we normalize the SFR and molecular gas masses by the surface area of the entire star forming disc of the host galaxies. Because the QSO positions are offset from the kinematic centre in a few cases, i.e. HE~1254$-$0934 and HE~1405$-$1545, we measure the size of the H$\alpha$ disc along the major-axis with respect to the kinematic centre assuming a circular disc. In Fig.~\ref{fig:Schmit_law}, we compare our measurements with the global Kennicutt-Schmidt law presented by \citet{Kennicutt:1998} and the relation for the spatially resolved law on kpc scales inferred from nearby star forming galaxies by \citet{Bigiel:2008}. Neither the disc-dominated nor the bulge-dominated QSO hosts are statistically different compared to the reference sample with the reasonably large scatter around the relation ($<$\,0.5\,dex). Only the ongoing major mergers are systematically above the Kennicutt-Schmidt relation, but also two elliptical galaxies with the highest SFR are located significantly above the relation with at least the same amplitude as the ongoing mergers. 

Overall, our combined optical IFU and single-dish \COzero\ observations seem to support a picture in which the cold gas in disc-dominated QSOs hosts are forming stars under similar conditions than normal star-forming disc galaxies. Apparently, there is no net effect by AGN feedback on the SFR in a positive nor negative sense. The bulge-dominated QSO host galaxies show more diverse properties. Despite the small sample size we find indications that the current SFR is enhanced and the gas depletion time is shorter in bulge-dominated QSO hosts compared to non-AGN galaxy sample with the same morphology.

\subsection{Molecular CO line temperature brightness ratios}
\begin{table}
\caption{CO(2-1)/CO(1-0) emission line ratio and aperture correction}
\label{tab:line_ratio}
\input{sample_ratio.tex}
\end{table}
For 7 of our targets we have secure detections of both the \COzero\ and \COtwo\ lines. The CO line ratio may provide first clues on the excitation conditions of the molecular gas. However, one issue is that the beam sizes are significantly different for \COzero\ ($22\arcsec$) and \COtwo\ ($11\arcsec$). Given the size of our galaxies, we require an aperture correction to provide conclusive results on the line ratios. Usually this correction is highly uncertain, but we showed in Section~\ref{sect:kinematics} that the H$\alpha$ and \COzero\ kinematics match. Thus, we use the measured H$\alpha$ flux distribution to estimate how much flux would be seen by the two different beam sizes assuming a purely 2D Gaussian beam. The estimated scaling factor $R_\mathrm{beam}$, representing the ratio of flux recovered with the $11\arcsec$ beam divided by the $22\arcsec$ beam. Here, we use the approach of down-scaling the measured $I_\mathrm{CO10}$ fluxes to the smaller beam of the \COtwo\ measurements by multiplying it with $R_\mathrm{beam}$. We then compute the temperature brightness ratio as
\begin{equation}
 r_{J\,J-1} =  I_{\mathrm{CO}J\,(J-1)}/I_{\mathrm{CO}(J-1)\,(J-2)} \times (J - 1)^2/J^2 \,.
\end{equation}
The initial flux measurement in units of mJy\,km/s, the beam correction factors and the resulting $r_{21}$ temperature brightness ratios are listed in Table~\ref{tab:line_ratio}.

We measure brightness temperature ratios in the range of 0.5-1.3 for our sample with a mean value of $\langle r_{21}\rangle=0.9\pm0.3$. This ratio is consistent with the range observed in normal star forming galaxies at low \citet{Braine:1992} and high-redshifts \citep[e.g.][]{Daddi:2015} as well as (ultra)-luminous infrared galaxies ((U)LIRGs, e.g. \citet{Papadopoulos:2012}). We note that the ratio is highest for the spiral galaxies, HE~1043$-$1346, HE~1237$-$2252 and HE~1239$-$2426, with extended spiral arms. For all other galaxies we measure a relatively low temperature brightness ratio around $r_{21}\sim0.5$ which might suggest that the CO emission is already sub-thermal at $J=2$. However, the temperature difference between \COzero\ and \COtwo\ is only 5\,K and the low-J line ratio is degenerate in terms of density and temperature in general. Furthermore, the spiral galaxies show a significantly higher SFR than the other galaxies except HE1254$-$0934 which is an ongoing gas-rich major merger.  Hence, it is difficult to draw robust conclusions just from these two low-J lines. Particularly, if the excitation condition of the CO have a radial gradient we would necessarily expect a significant difference given the range in the radial molecular gas distribution across the sample. Only high-J CO line observation would be able to robustly constrain the molecular gas excitation properties \citep[e.g.][]{Papadopoulos:2012,Indriolo:2017}.

\subsection{Comparing BH and host galaxy properties}
Various recent studies have investigated the link between the mass accretion rate of active BH and the SFR of their hosts. The AGN bolometric luminosity is a direct proxy of the mass accretion rate on the BH , $L_\mathrm{bol} = \eta c^2 \dot M_\mathrm{BH}$, for which a radiation efficiency $\eta$ of 10\,\% is usually assumed for AGN. Although the spatial and temporal scales of BH and star formation activity are orders of magnitude different, some studies reported a significant correlation between the integrated SFR and the BH mass accretion rate of AGN \citep[e.g.][]{Mullaney:2012,Chen:2013,Harris:2016}. However, other studies claimed that there is no significant correlation on host galaxy scales at all \citep[e.g.][]{Rosario:2012,Stanley:2015} or only for star formation on circumnuclear scales \citep[e.g.][]{Diamond-Stanic:2012}. \citet{Volonteri:2015} showed, based on detailed simulations of BH accretion in galaxies, that there is a strong difference in the results whether the sample is selected on SFR or on BH accretion rate. With our data we can go one step further by directly comparing the properties of the active BH with the molecular gas content and its distribution probed by the H$\alpha$ line paying attention to the host galaxy morphology.  

We characterize the active BHs in our QSO host galaxies with three basic parameters, BH mass ($M_\mathrm{BH}$), bolometric QSO luminosity ($L_\mathrm{bol}$), Eddington ratio ($L_\mathrm{bol}/L_\mathrm{Edd}$). Those are determined from the single-epoch QSO spectra using the rest-frame continuum luminosity at 5100$\AA$\ ($L_{5100}$) and the width of the broad H$\beta$ line adopting empirically derived scaling relations. We perform a multi-Gaussian fitting of the various broad and narrow-emission lines in the wavelength region containing the H$\beta$ and [OIII] lines above a local continuum as described in \citet{Husemann:2013a}. Based on the best-fit model we determine the QSO continuum flux at 5100\,\AA\ and the FWHM and line dispersion of the broad H$\beta$ line without the contribution of the narrow-line component on top of it. From those parameters we compute the bolometric luminosity as $L_\mathrm{bol} = 10\times L_{5100}$ which is consistent with empirically determined scaling relations between continuum luminosity and bolometric luminosity \citep[e.g.][]{Richards:2006}. $M_\mathrm{BH}$ can be calculated by adopting the virial theorem for the BLR clouds in combination with an empirical relation between continuum luminosity at 5100\,\AA\ and BLR size. Various different calibrations for virial $M_\mathrm{BH}$ estimates have been used in the literature that differ in the virial scale factor $f$, which takes into account the geometry and kinematic structure of the BLR, and the power-law slope $\alpha$ of the BLR size--luminosity relation. Here we adopt the calibration used by \citet{Assef:2011}
\begin{equation}
M_\mathrm{BH}=10^{6.83}f\left(\frac{\mathrm{FWHM}_{\mathrm{H}\beta}}{1000\,\mathrm{\mathrm{km}\,\mathrm{s}^{-1}}}\right)^2\left(\frac{L_{5100}}{10^{44}\,\mathrm{erg}\,\mathrm{s}^{-1}}\right)^{\alpha}\
 M_ { \sun }
\ ,
\end{equation}
with $\alpha=0.52$ \citep{Bentz:2009a} and $f=1.17$ following \citet{Collin:2006} for the FWHM as a surrogate for the kinematics of the BLR clouds. All directly measured quantities from the spectra and the derived BH parameters are tabulated in Table~\ref{tab:BH_parameters}. Here we list the results for the entire QSO sample presented in Paper I for completeness even though we only discuss a slightly smaller sub sample in this paper. 

\begin{table*}
 \caption{Black Hole mass properties for the entire QSO sample of Paper I.}\label{tab:BH_parameters}
 \input{sample_BH_results.tex}
\end{table*}

Since the QSO continuum luminosity and width of the broad H$\beta$ line can be determined from long-slit spectroscopy and do not rely on IFU observation, we also collect similar measurements for low-redshift QSOs from the literature. We focus on the unobscured QSOs observed in \COzero\ by \citet{Evans:2001}, \citet{Scoville:2003}, \citet{Evans:2006}, \citet{Bertram:2007}, and \citet{Evans:2009} that were primarily drawn from the Palomar Green survey \citep{Schmidt:1983} and Hamburg-ESO survey \citep{Wisotzki:2000}. In addition, we require that a rough morphological classification is available from deep imaging \citep{Guyon:2006, Kim:2008, Busch:2014}. Continuum luminosities and broad H$\beta$ line FWHM are taken from \citet{Boroson:1992}, \citet{Kaspi:2000}, \citet{Baskin:2005b}, \citet{Husemann:2013a}, and from \citet{Schulze:2010}. Since measurement errors are often not provided, we assume a canonical systematic error on the BH mass estimates of 0.3\,dex \citep[e.g.][]{Denney:2009}. 

\begin{figure}
\resizebox{\hsize}{!}{\includegraphics{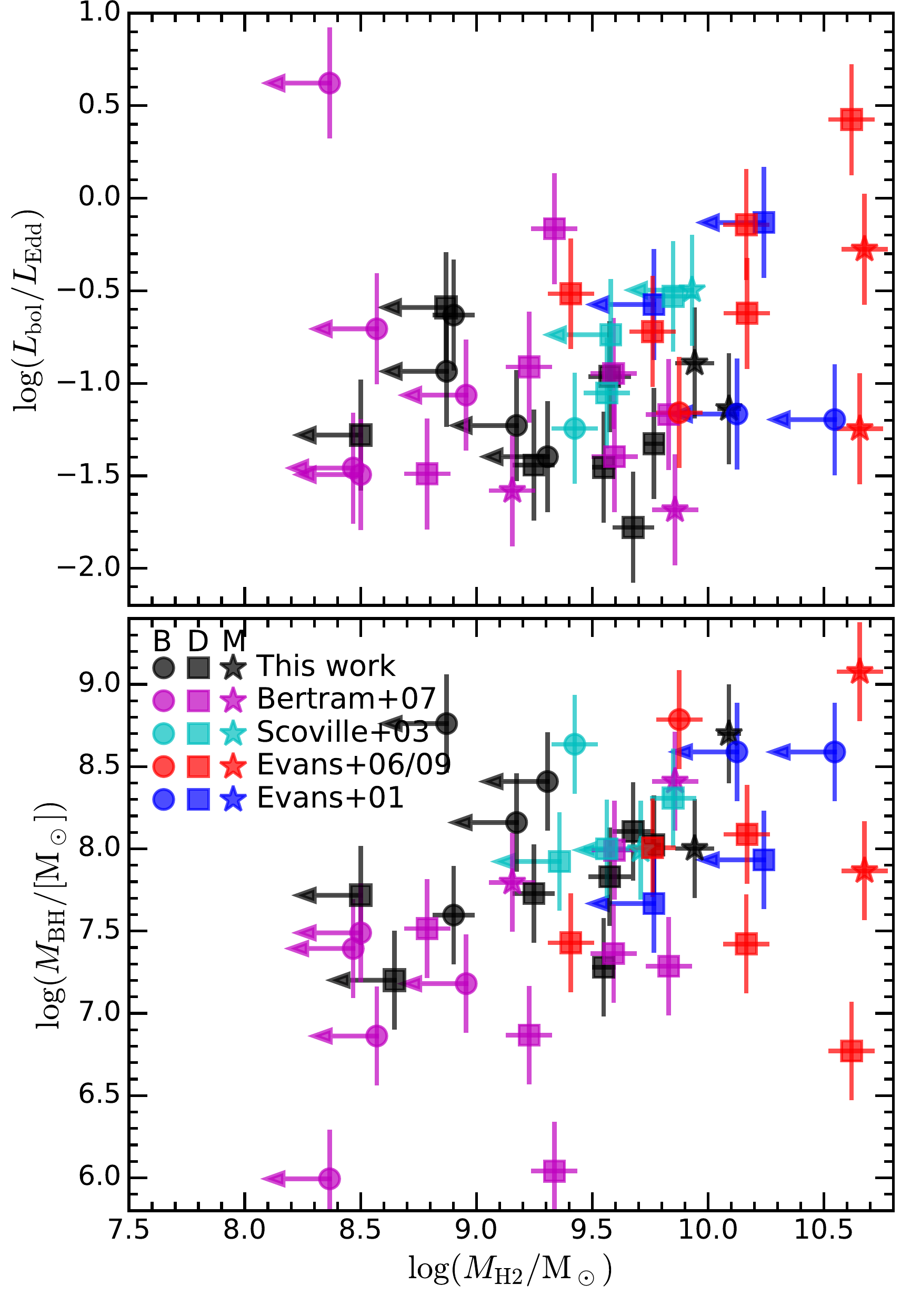}}
\caption{Comparison of Eddington ratio (upper panel) and black hole mass (lower panel) against the molecular gas content. The different symbols correspond to bulge-dominated (B), disc-dominated (D) and ongoing major merger (M) QSO host galaxies, while the symbol colour indicates different QSO samples from the literature.}
\label{fig:QSO_MBH_Edd}

\end{figure}

In Fig.~\ref{fig:QSO_MBH_Edd} we compare the Eddington ratio and the BH mass with molecular gas content of our QSO sample together with those taken from the literature. We find that the Eddington ratio is uncorrelated with the total molecular gas mass with a Kendall's tau correlation coefficient of 0.19. The mean value and rms scatter for the Eddington ratio of the combined sample is $\langle\log(L_\mathrm{bol}/L_\mathrm{Edd})\rangle=-0.9\pm0.5$. The correlation with the BH mass is stronger than with $L_\mathrm{bol}/L_\mathrm{Edd}$ given a Kendall's tau correlation coefficient of 0.35 and a 3\% chance that the data are uncorrelated.
\begin{figure*}
\centering
\includegraphics[width=\textwidth]{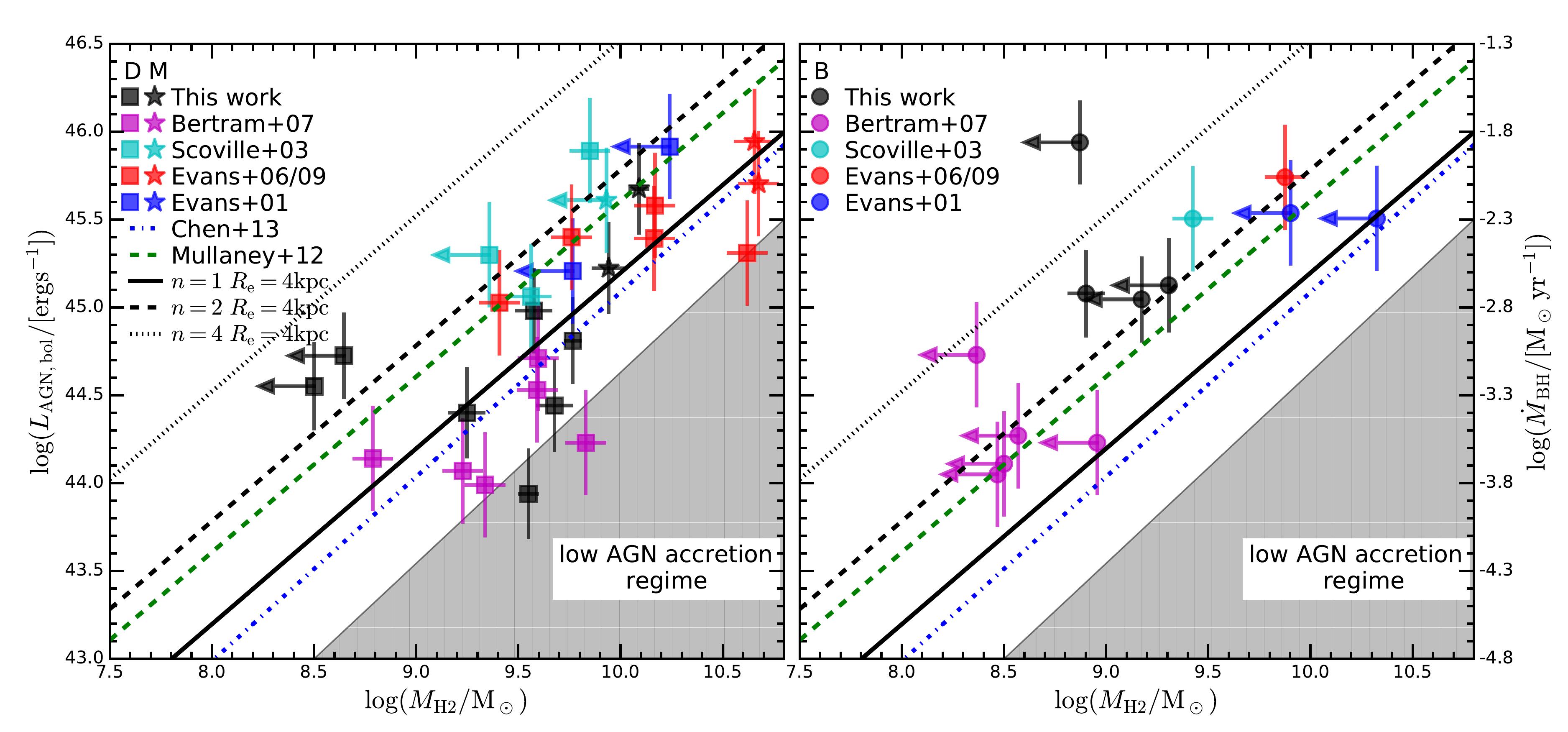}
\caption{Comparison of AGN bolometric luminosity or BH mass accretion rate against the molecular gas content for disc-dominated and major merger host galaxies (left panel) and bulge-dominated host galaxies (right panel). The different symbols and colours are consistent with Fig.~\ref{fig:QSO_MBH_Edd}. The dashed green and dashed-doted blue line correspond to the SFR-$L_\mathrm{bol}$ correlation reported by \citet{Mullaney:2012} and \citet{Chen:2013} assuming a gas depletion time scale of $10^9\,\mathrm{yr}$ typical for galaxies on the star-forming main sequence. The black line correspond to a simple model where the radial gas distribution is following a Sersi\'c-like profile with Sersi\'c index $n$ and an effective radius of $4$\,kpc of which only the inner 100\,pc is fueling the BH over a $10^7$\,yr. We also roughly indicate the region in which we certainly lack objects due to the flux-limited AGN selection as the grey shaded area. The area is just approximate, but highlights that the data points do not represent a true correlation but rather an upper envelope.}
\label{fig:QSO_Lum}
\end{figure*}

The relation between AGN luminosity and SFR has been investigated in many previous studies \citep[e.g.][]{Netzer:2009,Rosario:2012,Mullaney:2012,Chen:2013}. Here, we compare the AGN luminosity with the cold molecular gas content which is the actual driver for star formation and might also trace the reservoir of gas available for accretion onto the BH. In Fig.~\ref{fig:QSO_Lum} we compare the QSO bolometric luminosity against the total molecular gas mass for disc-dominated and merger hosts (left panel) and bulge-dominated hosts (right panel). We find a significant correlation between the BH accretion rate and the total gas content for the combined disc-dominated and merger host sample with Kendall's tau correlation coefficient of 0.45 and a probability that the two quantities are uncorrelated of $<$\,0.01\%. Only two disc-dominated QSOs with apparently low molecular gas content appear as strong outliers from the overall trend.  Interestingly, no statistically significant correlation is found for the bulge-dominated QSO host galaxies at all. The nature of the correlation for the disc-dominated and merger QSO host sample is surprising because it is unclear why the accretion happening on very small scale should be linked to the gas mass of the entire galaxy.

The first thing we want to emphasize is that we use a flux-limited QSO sample that consists only of the brightest QSOs. This is supported by the Eddington ratio distribution of our sample implying that most of our QSOs accrete at 10\% of the Eddington limit. However, the AGN luminosity varies on timescales of less then $10^7$\,yr which is much smaller than the gas depletion time scale for normal star forming galaxies. Thus, most massive galaxies with high molecular gas content are currently not in a bright AGN phase with a luminosity below the sample selection. Hence it is a sample selection effect that we miss galaxies with high molecular gas masses and low AGN luminosities. We would expect a weak or no correlation if we had selected a small sample of galaxies based on host galaxy properties, e.g. stellar mass, because we only rely on instantaneous AGN luminosities instead of a long-term average \citep[see][for a discussion on this issue]{Hickox:2014}. 

Nevertheless, we know from the primary AGN sample selection that we capture AGN accreting close to the Eddington limit, so that the QSO sample studied here is representative for all high luminosity AGN. Since we are not selecting AGN based on their host galaxy properties we cover several luminous AGN with a low gas content although we could only observe a few systems with low upper limits in the SFR for the \COzero\ follow-up given the long exposure times. Strikingly, we detect those kind of AGN hosts almost exclusively in bulge-dominated QSOs for which have a lower gas fraction as shown previously, but not among the disc-dominated systems. The clear trend recovered for disc-dominated QSO hosts means that either the accretion rate is already saturated, which is unlikely given that most AGN are still below the Eddington limit, or that the accretion is limited by the gas reservoir provided by the galaxy. In the latter case we would interpret the trend as an upper boundary of the AGN fueling efficiency of a galaxy rather than a proper correlation. 

Although we are not able to discriminate those two scenarios, we want to explore a physical origin of the relation between BH accretion rate and total SFR or gas mass which should be tightly linked.   Here, we test whether a well-defined relation between the total gas content and its surface brightness distribution can account for our observations with reasonable parameters. We assume a Sersi\'c profile for the radial surface brightness distribution with Sersi\'c index $n$ and effective radius $R_e$ as guided by the stellar light distribution of galaxies where $n$ ranges between 1 and 4. We adopt an effective radius of 4\,kpc which is a realistic value for the galaxies we have in our QSO sample despite some scatter. In order to convert the gas surface density into a time-averaged AGN luminosity we assume that the AGN will eat up all the gas within $<$\,100\,pc constantly during a duty cycle of $\tau_\mathrm{AGN}\sim10^7$\,yr with a standard radiation efficiency of $\eta=0.1$. We show the corresponding relation between gas mass and AGN luminosity for $n=1$, $n=2$ and $n=4$ in Fig.~\ref{fig:QSO_Lum}. Such a simple model with $1<n<2$ is indeed consistent with the observations and does not require very unrealistic numbers, although many assumed parameters are not well constrained and a significant scatter for the parameters of individual systems likely exists. The relations we predict 
in this way are close in the slope and absolute normalization to the AGN luminosity--SFR relations reported by \citet{Mullaney:2012} and \citet{Chen:2013} if we simply assume a gas depletion time scale of 1\,Gyr to convert the SFR to molecular gas content. However, the relation of \citet{Rosario:2012} for low redshift objects is significantly offset toward higher relative gas masses, which is likely caused by the different selection on SFR and not on AGN luminosity.

Interestingly, the bulge-dominated  QSO host galaxies are able to reach a significantly higher BH accretion rate at a given molecular gas mass compared to their disc-dominated counterparts. We verify this by computing a difference in gas mass $\Delta M_\mathrm{H2}$ with respect to the gas mass expected from $L_\mathrm{bol}$ based on our toy model with $n=1$ as described above. Comparing the distributions of $\Delta M_\mathrm{H2}$ including the censored data shows that both samples are not drawn from the same parent population at $>$95\% confidence. This significant difference may be caused by (a) the expulsion of a significant fraction of molecular gas in bulge-dominated QSO hosts due to AGN feedback, (b) the heating of gas to higher average gas temperatures which reduces the ability of the ISM to form molecular gas, or (c) a steeper radial surface gas density distribution or much more compact gas discs \citep{Davis:2013} in bulge-dominated galaxies which naturally leads to higher ratio of circum-nuclear to total gas content. With the current data we cannot discriminate between those scenarios and even a combination of those effects is possible.

\section{Discussion}\label{sect:discussion}
\subsection{The cold molecular gas content and the star formation efficiency in luminous QSOs}
If luminous AGN are able to drive large-scale gas outflow to quench star formation, a significant lower molecular gas content would be expected compared to non-AGN galaxies.
Only few studies so far have systematically investigated the cold gas content and the star formation efficiency in AGN host galaxies as the primary driver for star formation.  The molecular gas content of various small samples of low-redshift QSOs have been observed over the years \citep[e.g.][]{Scoville:2003,Bertram:2007,Villar-Martin:2013}, which generally turned out to be rich  in molecular gas as expected for non-AGN disc-dominated galaxies. \citet{Ho:2008} and \citet{Fabello:2011} compared the total HI atomic gas content of nearby AGN host galaxies and non-AGN galaxies and also reported no significant difference. A different result at very low redshift $z<0.03$ was presented by \citet{Saintonge:2012} based on the COLD~GASS survey. They found that AGN host galaxies have a lower molecular gas content than a matched control sample of non-AGN galaxies. However, a potential issue in the interpretation is caused by the fact that their AGN sample is dominated by Low-Ionisation Nuclear Emission Regions (LINERs) in which the emission maybe pre-dominantly powered by post-AGB stars rather than low-luminosity AGN \citep[e.g.][]{Singh:2013}. 

Observations of molecular gas have been difficult in luminous QSOs at $z>1$, but became possible with NOEMA and ALMA. \citet{Brusa:2015b} reported a significantly lower molecular gas content for a single QSOs with a strong ionized gas outflow. Extending these observations  to a larger sample of 11 QSOs with ALMA seems to confirm this trend \citep{Kakkad:2017}. The lower gas content also implies a higher star formation efficiency based on the FIR luminosity of the host galaxies, which may be caused by positive feedback of the AGN or by a different timescale of the star formation and AGN phase that wash out any causal connection.

In this work, we present new molecular gas observations for a low-redshift sample of QSOs. The advantage of this sample is that star formation rates can be inferred based on H$\alpha$ using optical integral-field spectroscopy in a spatially-resolved fashion (see Paper I for details).  Furthermore, we consider different morphological types provided by the deep optical imaging \citep{Jahnke:2004b}, which reveal an important indication of the evolutionary history of the systems. By separating the galaxies in disc-dominated, ongoing major mergers and bulge-dominated host galaxies, we can study the properties of pre-merging, merging and post-merging AGN host galaxies separately. 

For our disc-dominated QSO hosts we find that the gas fractions and depletion times are fully consistent with that of non-AGN star forming galaxies and follow the global Kennicutt-Schmidt relation with just two exceptions. This is in agreement with previous works at low redshift that the majority of luminous QSOs reside in gas-rich galaxies. Thus, AGN are apparently not able to efficiently remove the cold gas nor change the star formation efficiency in these host galaxies. This may be unsurprising because the AGN radiation is not isotropic in the framework of the AGN unification scheme and the released energy will more easily accelerate the low density gas along the galaxy poles. Recent high-resolution numerical simulation of the AGN feedback in disc-dominated systems found that the AGN had little impact \citep[e.g.][]{Gabor:2014} consistent with our observational data. The data for the two disc-dominated QSOs exhibit an exceptionally low molecular gas content. In these cases, we are not yet able to causally link the low gas content directly to AGN feedback event without resolving the outflow itself.

A high gas fraction and shorter depletion time, hence higher star formation efficiency, is observed for our ongoing major mergers. In both cases we find the star formation to be concentrated in a few regions, which leads to high gas surface densities. This is expected from numerical simulations \citep[e.g.][]{Mihos:1996,DiMatteo:2007,Hopkins:2013} and observed in many starburst galaxies with and without AGN \citep[e.g.][]{LiuX:2012,Schmidt:2013,Davies:2015}. It is interesting to note in this respect that the major merger system HE~1254$-$0934 shows two nuclei in continuum light separated by 8\,kpc containing the QSO and the starburst region, respectively. This naturally limits the AGN to influence the gas content just by geometrical considerations. It remains unclear whether the primary nucleus with the starburst also hosts a SMBH as expected during a major merger that could become active at later times to expel the cold gas in this region with some time delay.

Most interesting in this respect are the bulge-dominated QSO  host galaxies. They have been reported to exhibit significantly bluer colors than non-AGN counterparts \citep[e.g.][]{Jahnke:2004b,Sanchez:2004b,Zakamska:2006,Silverman:2008,Floyd:2013}. We confirm with the 3D spectroscopy that the exceptionally blue colors of the bulge-dominated systems \citep{Jahnke:2004b} are directly related to ongoing star formation placing those galaxies on the main-sequence of star formation in most cases (see Paper I). The  \COzero\ observations reported here usually lead to only upper limits except in one case, which implies that the depletion time scales are, on average, shorter compared to the disc-dominated galaxies. There are three plausible scenarios to explain the higher star formation efficiency in the bulge-dominated system:
\begin{itemize}
\item a recent remnant from a major merger with a remaining centrally-concentrated reservoir of gas 
\item re-rejuvenation with cold gas provided by one or several gas-rich minor mergers 
\item enhanced star formation efficiency caused by positive AGN feedback
\end{itemize}

The first scenario may be the most popular one. Assuming a gas rich major merger with $M_\mathrm{H2}>10^{10}M_{\sun}$ would still allow AGN feedback to expel most of the gas during the major merger with a remnant gas fraction that we would observe now after coalescence \citep[e.g.][]{Matteo:2005}. That the bulge-dominated systems are currently on the main-sequence of star formation may still be a significant reduction of the SFR if they had an enhanced SFR during the actual merger. However, one issue is that no obvious tidal features are usually seen in the available ground-based images. This is not an issue of depth since tidal features and/or minor companions are seen in disc-dominated systems, e.g. HE~1019$-$1414 or HE~1239$-$2426, as shown in \citet{Jahnke:2004b}. It still remains possible that even at the depth of 25.7\,mag/arcsec${^2}$ in $V$ band we may miss faint tidal features of a merger event about $\sim$1\,Gyr ago. Such a long time delay between the starburst and the AGN phase along the merger sequence \citep[e.g.][]{Wild:2010,Schawinski:2010b} could prevent the detection of signature from the past merger, but also means that any causal connection between the merger and current activity is obscure and ambiguous.

On the contrary, the molecular gas may be recently added through minor mergers that re-rejuvenated an already passive bulge-dominated galaxies as suggested by \citet{Hasinger:2008} or \citet{Khalatyan:2008}, to give two examples. In this scenario, no AGN feedback is needed as the SFR as well as the BH accretion is simply driven by the availability of the gas reservoir without the need for direct feedback. Since even quiescent bulge-dominated galaxies do contain a cold gas reservoir \citep{Young:2011}, the enhanced levels of star formation may be caused by positive AGN feedback where shocks are able to over-pressurize the gas considering the minor mergers allow may lead to suppressed star fromation rate \citep{Davis:2015}

\subsection{The link between BH and galaxy growth}
The connection between the level of BH accretion, corresponding to AGN luminosity, and galaxy growth through star formation has been a key area of AGN research. It is fundamentally linked to the nature of the fueling and feedback of AGN. The various studies reached quite different conclusions. For example, \citet{Netzer:2009b} reported a tight log-linear correlation at $z<0.1$ between AGN luminosity and SFR over several orders of magnitude based on the SDSS spectroscopic by including problematic LINERs as low-luminosity AGN. Similar linear trends over a more limited luminosity range were also reported by \citet{Mullaney:2012} and \citet{Chen:2013}. Probing the SFR of AGN host galaxies in various redshift bins out to $z<2.5$,  \citet{Rosario:2012} revealed a non-linear relation in the sense that the SFR becomes independent of AGN luminosity at low BH accretion rates. Several other studies also reported no correlation between AGN luminosity and SFR at all \citep{Azadi:2015,Stanley:2015,Shimizu:2016}. In the light of this unsolved issues we compared the molecular gas content as the primary driver for star formation with the BH accretion rate for our QSO sample and a collection of additional data of QSOs at redshift $z<0.3$ from the literature. 

We recover a positive correlation of molecular gas mass with BH accretion considering only disc-dominated and major merger host galaxies. This is qualitative in agreement with the work by \citet{Maiolino:2007}, but they combined samples up to redshift $z=5$ so that their result may be primarily driven by redshift and not by intrinsic accretion rate. Furthermore, one has to be careful to directly interpret the positive trend as proof for BH and galaxy co-evolution since \citet{Volonteri:2015} has shown that the results strongly depend on whether the sample is selected/binned in BH accretion or SFR (or equivalently molecular gas mass). Indeed, a lot of disc galaxies with a substantial gas reservoir are non-AGN galaxies with a very low BH accretion rate. This is a direct consequence of the AGN variability given the semi-stochastic process of BH accretion in which a circum-nuclear gas reservoir is a necessary but not sufficient requirement for BH accretion \citep[e.g.][]{Hickox:2014}. 

Since our galaxy sample is selected to host actively accreting BHs by design, the circum-nuclear gas reservoir may be considered the primary limitation of the BH accretion rate. Of course, bigger BHs can accrete more matter, given their Eddington limit, and artificially produce the trend found here. We can rule this out for our sample because we confirmed that any trend of molecular gas mass with Eddington ratio or BH mass is statistically much weaker. It still remains difficult to explain why the correlation between the BH accretion, potentially driven by the circum-nuclear gas mass, should be significantly correlated with the total gas mass of a galaxy on several kpc scales. With the aid of our toy model we show that invoking a global scaling relation between total and circum-nuclear gas mass is consistent with the observations for disc-dominated QSO host galaxies. It also reproduces previous empirical relations between SFR and BH accretion naturally adopting a depletion time scale of 1\,Gyr. 

Such a simple relation between total and circum-nuclear gas content breaks down for the bulge-dominated QSO host galaxies. Most of these systems in the sample have just upper limits for the molecular gas masses. Depending on their redshift, those limits are very stringent or slightly lower than that of the detected disc-dominated systems. This already indicates that their gas content must be systematically lower at a given AGN luminosity than their disc-dominated counterparts with high confidence. Considering cases such as HE~1029$-$1401 at $z=0.086$ highlights that the molecular gas content can be orders of magnitude lower to produce the equivalent BH accretion rate as expected from the disc-dominated hosts. The apparently uncorrelated growth of the BH with total molecular gas mass in bulge-dominated systems may help to wash out any trend for the total AGN host galaxy population depending on the details of the sample selection. Thus the morphology of galaxies needs to be additionally considered when the relation between BH accretion rate and total SFR/gas content is studied for AGN host galaxies.

Still it remains open why the BH accretion rate is enhanced so much in bulge-dominated galaxies in the light of the overall molecular gas content. Invoking again a simple scaling relation between circumnuclear gas surface density and the gas distribution on host galaxy scales, naturally leads to a steeper radial surface density relation or, alternatively, a much more compact gas disc. We already find indications that the gas surface densities are indeed higher  for the bulge-dominated compared disc-dominated QSO host galaxies indirectly through the higher star formation efficiencies and shorter gas depletion times. Alternatively, it could also be indirect evidence for negative AGN feedback that either removed a large fraction of the outer molecular gas content or heated the gas is to much higher temperatures not observable in the cold gas phase anymore. 
The temperature brightness ratios of the CO(2-1)/CO(1-0) obtained for our sample are not providing direct evidence for a higher cold gas temperature, but the low-J line ratio is not a very sensitive measure of the global gas temperature. Hence, more sensitive and higher resolution observations would be necessary in particular for the bulge-dominated QSOs to verify the circumnuclear gas surface densities and excitation conditions in the future with ALMA. 

\section{Summary and conclusions}\label{sect:conclusion}
In this paper, we presented new \COzero\ and \COtwo\ observations for a sample of 14 luminous nearby QSOs host galaxies that we interpret in conjunction with the available optical integral-field observations. This data set significantly extends the sample of luminous unobscured QSOs with molecular gas mass estimates at low redshift. Our main results of this work can be summarized as follow:
\begin{itemize}
 \item We detect the \COzero\ emission-line in 8 of 14 targets ($\sim$60\%) with a clear preference for disc-dominated and major merger systems with a detection rate of nearly 80\%.
 \item The integrated \COzero\ emission-line matches the spatially-integrated H$\alpha$ line profile indicating that both are tracing star formation regions when averaged over kpc scales.
 \item We also detect \COtwo\ emission-line in 7 of 11 targets ($\sim$60\%). The corresponding $r_{21}$ brightness temperature ratio is close to 1 and ranges between 0.5 and 1.3. Some of the lowest ratios may suggest that the CO emission is already sub-thermal at J = 2 in these systems.
 \item While the disc-dominated QSO host galaxies exhibit similar gas fractions and gas depletion time-scales compared to non-AGN host galaxies, we find that bulge-dominated QSO hosts have systematically smaller gas fractions but shorter gas depletion times given their elevated star formation rates. The two major mergers in our sample show high gas fraction and rather short gas depletion times as expected from simulations for strongly interacting systems.
 \item We report a positive correlation between BH accretion rate and total molecular gas mass for disc-dominated QSO host galaxies after combining our sample with literature data. We interpret this as an upper limit of BH accretion rate given our initial selection for the most luminous QSO at the considered redshift. We can explain this trend using a simply toy model that links the circum-nuclear gas surface density with the total gas mass through a simple scaling relation for discs.
 \item The BH accretion rate in bulge-dominated QSOs host galaxies is significantly higher given their molecular gas content compared to disc-dominated QSOs hosts. This may be explained either by a more compact or steeper gas surface density gradient, removal of the outer envelope of gas through feedback or a significantly higher gas temperature leading to a different composition of the ISM different phases. While we did not measure a higher CO line excitation based on the $r_{21}$ temperature brightness ratio, this needs to be investigated with more excitation-sensitive observations.
\end{itemize}

Despite the relatively small sample size of this study, we can confirm that the molecular gas content and star formation efficiency at least in disc-dominated QSO hosts and major mergers is indistinguishable from the corresponding population of non-AGN galaxies. Negative AGN feedback that either reduces the star formation efficiency or significantly lowers the cold gas content through galactic outflows is an implausible scenario for the majority of QSO hosts studied in this work. This is unsurprising for disc-dominated galaxies because the coupling efficiency between the AGN and the disc is assumed to be small due to the specific geometry and strong density gradients \citep{Gabor:2014}. If the majority of luminous AGN are hosted by disc-dominated system as observations start to indicate \citep[e.g.][]{Schawinski:2011,Cisternas:2011,Schawinski:2012}, it is easy to argue that galaxies and BHs co-evolve together only limited by availability of gas to fuel without the need for feedback. This may then explain why the cosmological evolution of the SFR density and the BH accretion rate density is similar with redshift as argued by \cite{Hickox:2014}. This notion is further supported by the trend between total molecular gas mass and BH accretion rate for low-redshift disc-dominated QSO host galaxies, which we can explain by invoking a global scaling relation between total gas mass and circum-nuclear gas surface density for gas rich discs. 

The bulge-dominated QSOs are the most interesting class of objects in our sample. Since the red sequence of quiescent galaxies is dominated by bulge-dominated galaxies, feedback from BH are naturally expected to primarily act on those galaxies \citep[e.g.][]{Schawinski:2006}. Interestingly, our bulge-dominated QSOs indeed exhibit a lower molecular gas mass fraction than would be na\"ively expected if an outflow had wiped out most of the cold gas content. However, the galaxies are not quiescent, and often show specific SFRs consistent with disc-dominated galaxies which leads to shorter gas depletion times. This rules out the standard form of negative AGN feedback to quench star formation and rather points to positive AGN feedback in the current phase of activity. Nevertheless, we should not forget the relatively short life time of AGN and their potential stochastic activity \citep[e.g.][]{Hickox:2014,Schawinski:2015}. A shock front at the speed of 1000 km/s requires about $10^6$\,yr to reach out to 1\,kpc distance from the BH. This means that depending on the actual life time of AGN, we may only be able to directly measure the effect of AGN feedback and outflows when the AGN already turned off. In this scenario, we are observing the bulge-dominated QSOs still in the phase before they totally cease their star formation.

\section*{Acknowledgments}
We thank the anonymous referee for providing constructive and helpful comments on short timescales that helped to significantly improve the paper. We acknowledge help from T. Bertram for providing us with electronic data for objects HE~1310$-$1051 and HE~1338$-$1423 from his IRAM 30m observations in 2007. Also, we appreciate discussions with David Rosario and Leonard Burtscher. 
TAD acknowledges support from a Science and Technology Facilities Council Ernest Rutherford Fellowship and HD acknowledges financial support from the Spanish Ministry of Economy and Competitiveness (MINECO) under the 2014 Ram\'on y Cajal program MINECO RYC-2014-15686. 
The article is based on observations carried out with the IRAM Plateau de Bure Interferometer (program 210-13; PI: B. Husemann). IRAM is supported by INSU/CNRS (France), MPG (Germany) and IGN (Spain). Based on observations made with VIMOS integral field spectrograph mounted to the Melipal VLT telescope at ESO-Paranal Observatory (programs 072B-0550 and 083B-0801; PI: K. Jahnke). This work made use of the matplotlib package \citep{Hunter:2007}, the CosmoCalc tools \citep{Wright:2006}, and the NASA/IPAC Extragalactic Database (NED), which is operated by the Jet Propulsion Laboratory, California Institute of Technology, under contract with the National Aeronautics and Space Administration.

\bibliographystyle{mnras}
\bibliography{references}

\end{document}

%% file: obs_log.tex
\begin{tabular}{lcccccccccc}\hline\hline
Name    	&  $\alpha$(J2000)	&  $\delta$(J2000) 	&			night     & El.      & $\nu_{\mathrm{E090}}$ & $\nu_{\mathrm{E230}}$ & $t_\mathrm{exp}$ & pwv & rms & comments\\
		   &            	&	  		&	     & [deg]  & [GHz] & [GHz] & [min] & [mm] & [mk] &\\\hline
HE~0952$-$1552 (D) &	148.62325	&	-16.11422	& 2014-01-23 & 35--37 & 102.8 & 207.4 & 37 & 4--8 & 0.5 & \\
		   &			&			& 2014-01-27 & 35--37 & 102.8 & 207.4 & 50 & $\sim$3 & &clear \\
HE~1019$-$1414 (D) &	155.60291	&	-14.48277	& 2014-01-24 & 36--38 & 106.5 & 214.1 & 80 & 3--4 & 0.5 &clear, strong wind\\
		   &			&			& 2014-01-28 & 35--38 & 106.8 & 214.1 & 30 & $\sim$1 &  & clear, strong wind\\
HE~1029$-$1401 (B) &	157.97625	&	-14.28083	& 2014-01-26 & 33--39 & 106.2 & 213.1 & 120 & 1.5--2 & 0.4 & clear \\
		   &			&			& 2014-01-27 & 37--39 & 106.8 & 214.1 & 50 & $\sim$1 & & clear \\
HE~1043$-$1346 (D) &	161.57083	&	-14.04111	& 2014-01-24 & 35--37 & 106.5 & 214.1 & 40 & $\sim$4 & 0.8 & clear, strong wind \\
HE~1110$-$1910 (B) &	168.21208  	&       -19.43972	& 2014-01-27 & 29--33 & 102.8 & 207.4 & 73 & 3--4.5 &  0.5 & clear, turbulent atmosphere \\
HE~1228$-$1637 (B) &	187.85291  	&       -16.89750	& 2014-01-26 & 29--33 & 106.2 & 213.1 & 55 & $\sim$2 & 0.4 & clear \\
		   &			&			& 2014-01-27 & 35--36 & 102.8 & 207.4 & 38 & 4--6 & & clouds, turbulent atmosphere\\
		   &			&			& 2014-01-28 & 33--36 & 106.8 & 214.1 & 37 & 1 & & clear\\
HE~1237$-$2252 (D) &	190.11791  	&	-23.15722	& 2014-01-26 & $\sim27$ & 106.2 & 213.1 & 45 & $\sim$2 & 0.9 & clear \\
HE~1239$-$2426 (D) &	190.65500  	&	-24.71111	& 2014-01-24 & $\sim28$ & 106.5 & 214.1 & 40 & 4--5 & 1.1 & clear, strong wind \\
HE~1254$-$0934 (M) &	194.23708	&	-9.83777	& 2014-01-24 & 39--42 & 100.3 & 205.4 & 32 & 4--5 & 0.9 & clear, strong wind \\
HE~1300$-$1325 (B) &	195.84263	&	-13.69269	& 2014-01-26 & 27--33 & 106.5 & 220.0 & 23 & $\sim$2 & 1.5 & clear \\
HE~1405$-$1545 (M) &	212.10208	&	-15.99138	& 2014-01-24 & 32--36 & 100.3 & ... & 60 & 4--5 & 0.7 & clear, strong wind \\
		   &			&			& 2014-01-27 & 30--36 & 100.3 & ... & 41 & 9--20 & & clouds, turbulent atmosphere\\
HE~1416$-$1256 (B) &	214.76591	&	-13.17908	& 2014-01-25 & 35--38 & 102.8 & 207.4 & 42 & $\sim$8 & 0.6 & cloudy, high humidity \\
		   &			&			& 2014-01-27 & 36--40 & 102.8 & 207.4 & 54 & 3--6 & & clouds, turbulent atmosphere\\
\hline
\end{tabular}

%% file: sample_results.tex
\begin{tabular}{ccccccccccccc}\hline\hline\noalign{\smallskip}
& & & & &\multicolumn{4}{c}{${}^{12}\mathrm{CO}(1-0)$} & &\multicolumn{2}{c}{${}^{12}\mathrm{CO}(2-1)$}   \\\cline{6-9}\cline{11-12}\noalign{\smallskip}
Object & $z$ & $M_*$ & SFR & $\Delta v$ & $I_\mathrm{CO}$ & FHWM${}_\mathrm{CO}$ & $ L^\prime_\mathrm{CO}/10^{8}$ & M($\mathrm{H}_2$) & & $I_\mathrm{CO}$ & FWHM${}_\mathrm{CO}$   \\ \hline\noalign{\smallskip}
 & & [$M_\odot$] & [$M_\odot$/yr] & [$\mathrm{km\,s}^{-1}$] & [$\mathrm{K\,km\,s}^{-1}$] & [$\mathrm{km\,s}^{-1}$] & [$\mathrm{K\,km\,s}^{-1}\,\mathrm{pc}^{2}$] & [$10^9\,M_\odot$] & & [$K\,\mathrm{km\,s}^{-1}$] & [$\mathrm{km\,s}^{-1}$]  \\ \hline\noalign{\smallskip}
HE 0952$-$1552 & 0.11 & 11.2 & 1.4 & 50.0 & $0.9\pm 0.2$  & $497\pm 83$ & $8.7\pm 1.7$ & $3.8\pm 0.7$ & & $1.4\pm 0.4$ & $360\pm 119$ \\
HE 1019$-$1414 & 0.08 & 10.8 & 0.7 & 50.0 & $0.8\pm 0.2$  & $385\pm 87$ & $4.1\pm 0.8$ & $1.8\pm 0.3$ & & $1.2\pm 0.3$ & $374\pm 99$ \\
HE 1029$-$1401 & 0.09 & 11.1 & 1.7 & 50.0 & $<0.3$  & ... & $<1.7$ & $<0.7$ & & $<0.3$ & ...\\
HE 1043$-$1346 & 0.07 & 10.9 & 8.8 & 50.0 & $2.0\pm 0.3$  & $420\pm 70$ & $8.1\pm 1.1$ & $3.5\pm 0.5$ & & $5.4\pm 0.4$ & $438\pm 28$ \\
HE 1110$-$1910 & 0.11 & 10.8 & 0.4 & 50.0 & $<0.3$  & ... & $<3.1$ & $<1.3$ & & $<1.1$ & ...\\
HE 1228$-$1637 & 0.10 & 10.8 & 8.1 & 50.0 & $0.20\pm 0.04$  & $106^{*}$ & $1.7\pm 0.4$ & $0.8\pm 0.2$ & & $0.75\pm 0.03$ & $103\pm 4$ \\
HE 1237$-$2252 & 0.10 & 11.1 & 4.5 & 50.0 & $1.4\pm 0.3$  & $313\pm 121$ & $10.9\pm 2.3$ & $4.7\pm 1.0$ & & $3.8\pm 0.3$ & $302\pm 23$ \\
HE 1239$-$2426 & 0.08 & 11.1 & 12.1 & 50.0 & $2.4\pm 0.3$  & $291\pm 39$ & $13.4\pm 1.7$ & $5.8\pm 0.7$ & & $7.3\pm 0.6$ & $404\pm 37$ \\
HE 1254$-$0934 & 0.14 & 11.2 & 69.3 & 50.0 & $1.9\pm 0.3$  & $300\pm 34$ & $28.3\pm 4.3$ & $12.3\pm 1.9$ & & $3.6\pm 0.5$ & $268\pm 27$ \\
HE 1300$-$1325 & 0.05 & 10.8 & 5.5 & 50.0 & $<1.0$  & ... & $<2.0$ & $<0.9$ & & ... & ...\\
HE 1310$-$1051 & 0.03 & 10.2 & 1.3 & 50.0 & $<0.2$  & ... & $<0.2$ & $<0.1$ & & $<0.2$ & ...\\
HE 1338$-$1423 & 0.04 & 11.1 & 1.7 & 50.0 & $<0.2$  & ... & $<0.3$ & $<0.1$ & & ... & ...\\
HE 1405$-$1545 & 0.19 & 11.2 & 37.7 & 50.0 & $0.7\pm 0.2$  & $293\pm 69$ & $20.1\pm 4.4$ & $8.8\pm 1.9$ & & ... & ... \\
HE 1416$-$1256 & 0.13 & 10.4 & 2.2 & 50.0 & $<0.3$  & ... & $<4.3$ & $<1.8$ & & $<0.8$ & ...\\
\noalign{\smallskip}\hline\end{tabular}

%% file: sample_ratio.tex
\begin{tabular}{ccccc}\hline\hline
Object & $I_\mathrm{CO10}$ & $I_\mathrm{CO21}$ & $R_\mathrm{beam}$ & $r_{21}$ \\
 & [Jy\,km/s] & [Jy\,km/s] &  & \\\hline
HE 0952$-$1552 & $5.4\pm1.1$ & $10.6\pm4.1$ & 0.92 & $0.5\pm0.3$\\
HE 1019$-$1414 & $4.8\pm1.0$ & $8.8\pm2.0$ & 0.95 & $0.5\pm0.2$\\
HE 1043$-$1346 & $12.1\pm1.4$ & $40.4\pm3.7$ & 0.83 & $1.0\pm0.1$\\
HE 1228$-$1637 & $1.3\pm0.3$ & $5.6\pm0.4$ & 0.95 & $1.2\pm0.3$\\
HE 1237$-$2252 & $8.5\pm1.8$ & $28.2\pm3.1$ & 0.84 & $1.0\pm0.3$\\
HE 1239$-$2426 & $14.1\pm1.4$ & $54.6\pm3.7$ & 0.73 & $1.3\pm0.2$\\
HE 1254$-$0934 & $11.1\pm1.3$ & $27.3\pm3.4$ & 0.91 & $0.7\pm0.1$\\
\hline\end{tabular}

%% file: sample_BH_results.tex
\begin{tabular}{lcccccccccc}\hline\hline\noalign{\smallskip}
Object & $f_{5100}$ & $\log(L_{5100})$ & $\sigma_{\mathrm{H}\beta}$ & $\mathrm{FWHM}_{\mathrm{H}\beta}$ & $\log(M_\mathrm{BH})$ & $\log(L_\mathrm{bol})$ & $\log(L_\mathrm{bol}/L_\mathrm{Edd})$ \\ \noalign{\smallskip}
 & $\left[10^{-16}\frac{\mathrm{erg}}{\mathrm{s}\,\mathrm{cm}^2\,\textup{\AA}}\right]$ & $\left[\mathrm{erg\,s}^{-1}\right]$ & $\left[\mathrm{km\,s}^{-1}\right]$ & $\left[\mathrm{km\,s}^{-1}\right]$ & $[M_\odot]$ & $\left[\mathrm{erg\,s}^{-1}\right]$ &  \\ \hline\noalign{\smallskip}
HE~0952$-$1552 & $5.7\pm0.1$ & $44.0\pm0.1$ & $3517\pm 256$  & $2959\pm 143$  & $7.83\pm 0.07$ & $44.98\pm 0.24$ & $-1.62\pm 0.32$ \\
HE~1019$-$1414 & $3.2\pm0.2$ & $43.4\pm0.1$ & $1582\pm 106$  & $3726\pm 241$  & $7.73\pm 0.08$ & $44.40\pm 0.26$ & $-1.21\pm 0.34$ \\
HE~1029$-$1401 & $86.9\pm0.3$ & $44.9\pm0.1$ & $2066\pm 21$  & $4866\pm 50$  & $8.76\pm 0.05$ & $45.94\pm 0.24$ & $-0.70\pm 0.29$ \\
HE~1043$-$1346 & $1.4\pm0.1$ & $42.9\pm0.1$ & $5618\pm 358$  & $2928\pm 67$  & $7.28\pm 0.06$ & $43.94\pm 0.26$ & $-2.53\pm 0.34$ \\
HE~1110$-$1910 & $6.2\pm0.1$ & $44.0\pm0.1$ & $1989\pm 63$  & $4158\pm 198$  & $8.16\pm 0.06$ & $45.05\pm 0.25$ & $-1.09\pm 0.30$ \\
HE~1201$-$1409 & $7.3\pm0.1$ & $44.3\pm0.1$ & $1603\pm 102$  & $1794\pm 26$  & $7.57\pm 0.05$ & $45.32\pm 0.24$ & $-0.78\pm 0.32$ \\
HE~1228$-$1637 & $7.9\pm0.2$ & $44.1\pm0.1$ & $1244\pm 26$  & $2132\pm 17$  & $7.60\pm 0.05$ & $45.08\pm 0.25$ & $-0.67\pm 0.31$ \\
HE~1237$-$2252 & $2.1\pm0.1$ & $43.4\pm0.1$ & $2373\pm 81$  & $5613\pm 197$  & $8.11\pm 0.07$ & $44.44\pm 0.26$ & $-1.54\pm 0.33$ \\
HE~1239$-$2426 & $7.1\pm0.2$ & $43.8\pm0.1$ & $1738\pm 43$  & $4092\pm 109$  & $8.02\pm 0.05$ & $44.81\pm 0.25$ & $-1.09\pm 0.30$ \\
HE~1254$-$0934 & $15.8\pm0.9$ & $44.7\pm0.1$ & $2253\pm 40$  & $5306\pm 95$  & $8.70\pm 0.06$ & $45.68\pm 0.26$ & $-0.90\pm 0.32$ \\
HE~1300$-$1325 & $10.7\pm0.1$ & $43.5\pm0.1$ & $2373\pm 19$  & $5587\pm 43$  & $8.11\pm 0.05$ & $44.46\pm 0.24$ & $-1.53\pm 0.29$ \\
HE~1310$-$1051 & $24.9\pm0.8$ & $43.6\pm0.1$ & $1425\pm 61$  & $3359\pm 253$  & $7.72\pm 0.08$ & $44.55\pm 0.25$ & $-1.04\pm 0.32$ \\
HE~1315$-$1028 & $3.3\pm0.1$ & $43.7\pm0.1$ & $2087\pm 56$  & $4937\pm 62$  & $8.10\pm 0.05$ & $44.65\pm 0.25$ & $-1.33\pm 0.31$ \\
HE~1335$-$0847 & $17.5\pm0.8$ & $44.2\pm0.1$ & $3393\pm 542$  & $1455\pm 74$  & $7.32\pm 0.07$ & $45.18\pm 0.26$ & $-1.49\pm 0.41$ \\
HE~1338$-$1423 & $25.1\pm0.5$ & $43.7\pm0.1$ & $1120\pm 84$  & $1671\pm 63$  & $7.20\pm 0.06$ & $44.73\pm 0.25$ & $-0.75\pm 0.33$ \\
HE~1405$-$1545 & $2.6\pm0.1$ & $44.2\pm0.1$ & $1323\pm 17$  & $3114\pm 36$  & $8.00\pm 0.06$ & $45.22\pm 0.26$ & $-0.65\pm 0.32$ \\
HE~1416$-$1256 & $5.3\pm0.4$ & $44.1\pm0.1$ & $2242\pm 44$  & $5279\pm 126$  & $8.41\pm 0.07$ & $45.13\pm 0.27$ & $-1.16\pm 0.33$ \\
HE~1434$-$1600 & $22.7\pm0.4$ & $44.9\pm0.1$ & $2646\pm 38$  & $6230\pm 86$  & $8.94\pm 0.05$ & $45.87\pm 0.24$ & $-0.94\pm 0.29$ \\
\noalign{\smallskip}\hline\end{tabular}